\pdfoutput=1

\documentclass[a4paper,cits]{JINST}
\usepackage[version=3]{mhchem}
\usepackage[hyphens]{url}
\usepackage{booktabs}
\usepackage{bm}
\usepackage{microtype}

\let\ifpdf\relax 

\newcommand{\bb}{\ensuremath{\beta\beta}}
\newcommand{\bbonu}{\ensuremath{\beta\beta0\nu}}
\newcommand{\bbtnu}{\ensuremath{\beta\beta2\nu}}
\newcommand{\Qbb}{\ensuremath{Q_{\beta\beta}}}
\newcommand{\mbb}{\ensuremath{m_{\beta\beta}}}

\newcommand{\ckky}{\ensuremath{\mathrm{counts}/(\mathrm{keV} \cdot \mathrm{kg} \cdot \mathrm{y})}}

\newcommand{\XE}{\ensuremath{{}^{136}\mathrm{Xe}}}

\newcommand{\CS}{\ensuremath{{}^{137}\mathrm{Cs}}}
\newcommand{\GE}{\ensuremath{{}^{76}\rm Ge}}

\newcommand{\TL}{\ensuremath{{}^{208}\rm{Tl}}}

\newcommand{\BI}{\ensuremath{{}^{214}}Bi}

\newcommand{\Pb}{\ensuremath{^{208}}Pb}

\title{NEXT-100 Technical Design Report (TDR).\\ Executive Summary}

\author{
V.~\'Alvarez,$^{a}$
F.I.G.M.~Borges,$^{b}$
S.~C\'arcel,$^{a}$
J.M.~Carmona,$^{c}$
J.~Castel,$^{c}$
J.M.~Catal\'a,$^{d}$
S.~Cebri\'an,$^{c}$
A.~Cervera,$^{a}$
D.~Chan,$^{e}$
C.A.N.~Conde,$^{b}$
T.~Dafni,$^{c}$
T.H.V.T.~Dias,$^{b}$
J.~D\'iaz,$^{a}$
M.~Egorov,$^{e}$
R.~Esteve,$^{d}$
P.~Evtoukhovitch,$^{f}$
L.M.P.~Fernandes,$^{b}$
P.~Ferrario,$^{a}$
A.L.~Ferreira,$^{g}$
E.~Ferrer-Ribas,$^{h}$
E.D.C.~Freitas,$^{b}$
V.M.~Gehman,$^{e}$
A.~Gil,$^{a}$
I.~Giomataris,$^{h}$
A.~Goldschmidt,$^{e}$
H.~G\'omez,$^{c}$
J.J.~G\'omez-Cadenas,$^{a}$\thanks{Spokesperson (gomez@mail.cern.ch)}~
K.~Gonz\'alez,$^{a}$
D. Gonz\'alez-D\'iaz,$^{c}$
R.M.~Guti\'errez,$^{i}$
J.~Hauptman,$^{j}$
J.A.~Hernando Morata,$^{k}$
D.C.~Herrera,$^{c}$
V.~Herrero,$^{d}$
F.J.~Iguaz,$^{h}$
I.G.~Irastorza,$^{c}$
V.~Kalinnikov,$^{f}$
D.~Kiang,$^{e}$
L.~Labarga,$^{l}$
I.~Liubarsky,$^{a}$
J.A.M.~Lopes,$^{b}$
D.~Lorca,$^{a}$
M.~Losada,$^{i}$
G.~Luz\'on,$^{c}$
A.~Mar\'i,$^{d}$
J.~Mart\'in-Albo,$^{a}$
A.~Mart\'inez,$^{a}$
T.~Miller,$^{e}$
A.~Moiseenko,$^{f}$
F.~Monrabal,$^{a}$
C.M.B.~Monteiro,
J.M.~Monz\'o,$^{c}$
F.J.~Mora,$^{c}$
L.M. Moutinho,$^{g}$
J.~Mu\~noz Vidal,$^{a}$
H.~Natal da Luz,$^{b}$
G.~Navarro,$^{i}$
M.~Nebot,$^{a}$
D.~Nygren,$^{e}$
C.A.B.~Oliveira,$^{eg}$
R.~Palma,$^{m}$
J.~P\'erez,$^{n}$
J.L.~P\'erez Aparicio,$^{m}$
J.~Renner,$^{e}$
L.~Ripoll,$^{o}$
A.~Rodr\'iguez,$^{c}$
J.~Rodr\'iguez,$^{a}$
F.P.~Santos,$^{b}$
J.M.F.~dos Santos,$^{b}$
L.~Segui,$^{c}$
L.~Serra,$^{a}$
D.~Shuman,$^{e}$
C.~Sofka,$^{p}$
M.~Sorel,$^{a}$
J.F.~Toledo,$^{d}$
A.~Tom\'as,$^{c}$
J.~Torrent,$^{o}$
Z.~Tsamalaidze,$^{f}$
D.~V\'azquez,$^{k}$
E.~Velicheva,$^{f}$
J.F.C.A.~Veloso,$^{g}$
J.A.~Villar,$^{c}$
R.C.~Webb,$^{p}$
T.~Weber,$^{e}$
J.~White$^{p}$
and N.~Yahlali$^{a}$
\\
\llap{$^{a}$}
Instituto de F\'isica Corpuscular (IFIC), CSIC \& Universitat de Val\`encia\\
Calle Catedr\'atico Jos\'e Beltr\'an, 2, 46980 Paterna, Valencia, Spain\\
\llap{$^{b}$}
Departamento de Fisica, Universidade de Coimbra\\
Rua Larga, 3004-516 Coimbra, Portugal\\
\llap{$^c$}
Laboratorio de F\'isica Nuclear y Astropart\'iculas, Universidad de Zaragoza\\ 
Calle Pedro Cerbuna, 12, 50009 Zaragoza, Spain\\
\llap{$^d$}
Instituto de Instrumentaci\'on para Imagen Molecular (I3M), Universitat Polit\`ecnica de Val\`encia\\ 
Camino de Vera, s/n, Edificio 8B, 46022 Valencia, Spain\\
\llap{$^{e}$}
Lawrence Berkeley National Laboratory (LBNL)\\
1 Cyclotron Road, Berkeley, California 94720, USA\\
\llap{$^{f}$}
Joint Institute for Nuclear Research (JINR)\\
Joliot-Curie 6, 141980 Dubna, Russia\\
\llap{$^{g}$}Institute of Nanostructures, Nanomodelling and Nanofabrication (i3N), Universidade de Aveiro\\
Campus de Santiago, 3810-193 Aveiro, Portugal\\
\llap{$^{h}$}IRFU, Centre d'\'Etudes Nucl\'eaires de Saclay (CEA-Saclay)\\
91191 Gif-sur-Yvette, France\\
\llap{$^{i}$}
Centro de Investigaciones en Ciencias B\'asicas y Aplicadas, Universidad Antonio Nari\~no\\ 
Carretera 3 este No.\ 47A-15, Bogot\'a, Colombia\\
\llap{$^{j}$}
Department of Physics and Astronomy, Iowa State University\\
12 Physics Hall, Ames, Iowa 50011-3160, USA\\
\llap{$^{k}$}
Instituto Gallego de F\'isica de Altas Energ\'ias (IGFAE), Univ.\ de Santiago de Compostela\\
Campus sur, R\'ua Xos\'e Mar\'ia Su\'arez N\'u\~nez, s/n, 15782 Santiago de Compostela, Spain\\
\llap{$^{l}$}
Departamento de F\'isica Te\'orica, Universidad Aut\'onoma de Madrid\\
Campus de Cantoblanco, 28049 Madrid, Spain\\
\llap{$^{m}$}
Dpto.\ de Mec\'anica de Medios Continuos y Teor\'ia de Estructuras, Univ.\ Polit\`ecnica de Val\`encia\\
Camino de Vera, s/n, 46071 Valencia, Spain\\
\llap{$^{n}$}
Instituto de F\'isica Te\'orica (IFT), UAM/CSIC\\
Campus de Cantoblanco, 28049 Madrid, Spain\\
\llap{$^{o}$}
Escola Polit\`ecnica Superior, Universitat de Girona\\
Av.~Montilivi, s/n, 17071 Girona, Spain\\
\llap{$^{p}$}
Department of Physics and Astronomy, Texas A\&M University\\
College Station, Texas 77843-4242, USA\\
}

\abstract{
In this \emph{Technical Design Report} (TDR) we describe the NEXT-100 detector that will search for neutrinoless double beta decay (\bbonu) in \XE\ at the Laboratorio Subterr\'aneo de Canfranc (LSC), in Spain. The document formalizes the design presented in our \emph{Conceptual Design Report} (CDR): an electroluminescence time projection chamber, with separate readout planes for calorimetry and tracking, located, respectively, behind cathode and anode. The detector is designed to hold a maximum of about 150 kg of xenon at 15 bar, or 100 kg at 10 bar. This option builds in the capability to increase the total isotope mass by 50\% while keeping the operating pressure at a manageable level. 

The readout plane performing the energy measurement is composed of Hamamatsu R11410-10 photomultipliers, specially designed for operation in low-background, xenon-based detectors. Each individual PMT will be isolated from the gas by an individual, pressure resistant enclosure and will be coupled to the sensitive volume through a sapphire window. The tracking plane consists in an array of Hamamatsu S10362-11-050P MPPCs used as tracking pixels. They will be arranged in square boards holding 64 sensors ($8\times8$) with a 1-cm pitch. The inner walls of the TPC, the sapphire windows and the boards holding the MPPCs will be coated with tetraphenyl butadiene (TPB), a wavelength shifter, to improve the light collection.
}

\keywords{Time Projection Chambers (TPC)}
\preprint{arXiv:1202.0721}

\begin{document}

%%%%%%%%%%%%%%%%%%%%%%%%%%%%%%%%%%%%%%%%%%%%%%%%%%%%%%%%%%%%

\section{Introduction} \label{sec:Introduction}
Neutrinoless double beta decay (\bbonu) is a hypothetical, very slow nuclear transition in which two neutrons undergo $\beta$-decay simultaneously and without the emission of neutrinos. The importance of this process goes beyond its intrinsic interest: an unambiguous observation would establish that neutrinos are Majorana particles --- that is to say, truly neutral particles identical to their antiparticles --- and prove that total lepton number is not a conserved quantity in nature.

After 70 years of experimental effort, no compelling evidence for the existence of \bbonu\ has been obtained. A new generation of experiments that are already running or about to run promises to push forward the current limits exploring the degenerate region of neutrino masses (see \cite{GomezCadenas:2011it} for a recent review of the field). In order to do that, the experiments are using masses of $\bb$ isotope ranging from tens of kilograms to several hundreds, and will need to improve the background rates achieved by previous experiments by, at least, an order of magnitude. If no signal is found, masses in the ton scale and further background reduction will be required. Only a few of the new-generation experiments can possibly be extrapolated to those levels.

The \emph{Neutrino Experiment with a Xenon TPC} (NEXT) will search for neutrinoless double beta decay in \XE. A xenon gas time projection chamber offers scalability to large masses of \bb\ isotope and a background rate among the lowest predicted for the new generation of experiments \cite{GomezCadenas:2011it}, stemming from a radiopure setup and the tracking capabilities of the detector. The NEXT experiment was proposed to the \emph{Laboratorio Subterr\'aneo de Canfranc} (LSC), Spain, in 2009 \cite{Granena:2009it}, with a source mass of the order of 100 kg. Three years of intense R\&D have resulted in a \emph{Conceptual Design Report} (CDR) \cite{Alvarez:2011my} and a \emph{Technical Design Report} (TDR), summarized in this document, where the final design of the NEXT-100 detector is defined. More detailed reports on the design of the different subsystems will be forthcoming.

\section{Neutrinoless double beta decay searches} \label{sec:BBSearches}
Double beta decay (\bb) is a very rare nuclear transition in which a nucleus with $Z$ protons decays into a nucleus with $Z+2$ protons and same mass number $A$. It can only be observed in those isotopes where the decay to the $Z+1$ isobar is forbidden or highly suppressed. Two decay modes are usually considered:
\begin{itemize}
\item The standard two-neutrino mode (\bbtnu), consisting in two simultaneous beta decays, $\ce{^{A}_{Z}X} \to\ \ce{^{A}_{Z+2}Y} + 2\ e^{-} + 2\ \overline{\nu}_{e}$, which has been observed in several isotopes with typical half-lives in the range of $10^{18}$--$10^{21}$ years (see, for instance, \cite{GomezCadenas:2011it} and references therein). 
\item The neutrinoless mode (\bbonu), $\ce{^{A}_{Z}X} \rightarrow \ce{^{A}_{Z+2}Y} + 2\ e^{-}$, which violates lepton-number conservation, and is therefore forbidden in the Standard Model of particle physics. An observation of \bbonu\ would prove that neutrinos are massive, Majorana particles \cite{Schechter:1981bd}. The Heidelberg-Moscow experiment set the most sensitive limit to the half-life of \bbonu\ so far: $T^{0\nu}_{1/2}(\GE) \ge 1.9\times10^{25}$ years (90\% CL) \cite{KlapdorKleingrothaus:2000sn}. In addition, a subgroup of the experiment observed evidence of a positive signal, with a best value for the half-life of $1.5\times10^{25}$ years \cite{KlapdorKleingrothaus:2001ke}. The claim was very controversial \cite{Aalseth:2002dt}, and still awaits an experimental response.
\end{itemize}

The implications of experimentally establishing the existence of \bbonu\ would be profound. First, it would demonstrate that total lepton number is violated in physical phenomena, an observation that could be linked to the cosmic asymmetry between matter and antimatter through the process known as \emph{leptogenesis} \cite{Fukugita:1986hr, Davidson:2008bu}. Second, a Majorana nature for the neutrino, in contrast to the Dirac nature of the other Standard Model fermions, may explain the smallness of neutrino masses \cite{GomezCadenas:2011it}.

Several underlying mechanisms --- involving, in general, physics beyond the Standard Model  --- have been proposed for \bbonu, the simplest one being the virtual exchange of light Majorana neutrinos. Assuming this to be the dominant one at low energies, the half-life of \bbonu\ can be written as
\begin{equation}
(T^{0\nu}_{1/2})^{-1} = G^{0\nu} \ \big|M^{0\nu}\big|^{2} \ \mbb^{2}.
\end{equation}
In this equation, $G^{0\nu}$ is an exactly-calculable phase-space integral for the emission of two electrons; $M^{0\nu}$ is the nuclear matrix element of the transition, that has to be evaluated theoretically; and \mbb\ is the \emph{effective Majorana mass} of the electron neutrino:
\begin{equation}
\mbb = \Big| \sum_{i} U^{2}_{ei} \ m_{i} \Big| \ ,
\end{equation}
where $m_{i}$ are the neutrino mass eigenstates and $U_{ei}$ are elements of the neutrino mixing matrix. Therefore a measurement of the \bbonu\ decay rate would provide direct information on neutrino masses \cite{GomezCadenas:2011it}.

The detectors used in double beta decay experiments are designed to measure the energy of the radiation emitted by a \bb\ source. In the case of \bbonu, the sum of the kinetic energies of the two released electrons is always the same, and corresponds to the mass difference between the parent and the daughter nuclei: $\Qbb \equiv M(Z,A)-M(Z+2,A)$. However, due to the finite energy resolution of any detector, \bbonu\ events are reconstructed within a non-zero energy range centered around \Qbb, typically following a gaussian distribution. Other processes occurring in the detector can fall in that region of energies, thus becoming a background and compromising drastically the experiment's expected sensitivity to \mbb\ \cite{GomezCadenas:2010gs}.

All double beta decay experiments have to deal with an intrinsic background, the \bbtnu, that can only be suppressed by means of good energy resolution. Backgrounds of cosmogenic origin force the underground operation of the detectors. Natural radioactivity emanating from the detector materials and surroundings can easily overwhelm the signal peak, and consequently careful selection of radiopure materials is essential. Additional experimental signatures that allow the distinction of signal and background are a bonus to provide a robust result.

\section{The NEXT concept} \label{sec:Next}
New-generation double beta decay experiments have to be sensitive to lifetimes longer than $10^{25}$ years (or, equivalently, effective Majorana neutrino masses smaller than 100 meV). Designing a detector capable of identifying efficiently and unambiguously such a rare signal is a major experimental challenge. Consequently, many different techniques have been proposed, each one with its pros and cons. In order to compare them, a figure of merit, the sensitivity to \mbb, is typically used \cite{GomezCadenas:2010gs}:
\begin{equation}
\mbb \propto \sqrt{1/\varepsilon}\ \left( \frac{b\ \Delta E}{M\ t} \right)^{1/4} ,
\end{equation}
where $\varepsilon$ is the signal detection efficiency, $M$ is the \bb\ isotope mass used in the experiment, $t$ is the data-taking time, $\Delta E$ is the energy resolution and $b$ is the background rate in the region of interest around \Qbb\ (usually expressed in terms of counts per kg of \bb\ isotope, year and keV).

Some experiments, such as the germanium calorimeters or the bolometers, emphasize the energy resolution and the detection efficiency; other, like the separate-source trackers, use event kinematical information to discriminate signal and background, thus obtaining very low background rates. The NEXT experiment combines good energy resolution, a low background rate  and the possibility to scale-up the detector to large masses of \bb\ isotope by using a high-pressure xenon gas (HPXe) electroluminescent time projection chamber (TPC) to search for \bbonu\ in \XE. The combination results in excellent sensitivity to \mbb. For a total exposure of 500 ${\rm kg}\cdot${\rm year}, the sensitivity is better than 100 meV. This sensitivity can match or outperform that of the best experiments in the field \cite{GomezCadenas:2010gs}. 

Xenon is a suitable detection medium that provides both scintillation and ionization signals. In its gaseous phase, xenon can provide high energy resolution, better than 0.5\% at 2500 keV \cite{Nygren:2009zz}. Two naturally-occurring isotopes of xenon can decay \bb, $^{134}$Xe ($\Qbb = 825$ keV) and $^{136}$Xe ($\Qbb = 2458$ keV). The latter, having a higher $Q$-value, is preferred for neutrinoless double beta decay searches because the decay rate is proportional to $\Qbb^{5}$ and the radioactive backgrounds are less abundant at higher energies. The two-neutrino decay mode of \XE\ is slow, $\sim2.3\times10^{21}$ years \cite{Ackerman:2011gz,Gando:2012kz}, and hence the experimental requirement for good energy resolution is less stringent than for other \bb\ sources. \XE\ constitutes 8.86\% of all natural xenon, but the enrichment process is relatively simple and cheap compared to that of other \bb\ isotopes, thus making \XE\ the most obvious candidate for a future multi-ton experiment. Also, xenon, unlike other \bb\ sources, has no long-lived radioactive isotopes that could become a background.

\begin{figure}
\centering
\includegraphics[width=8cm]{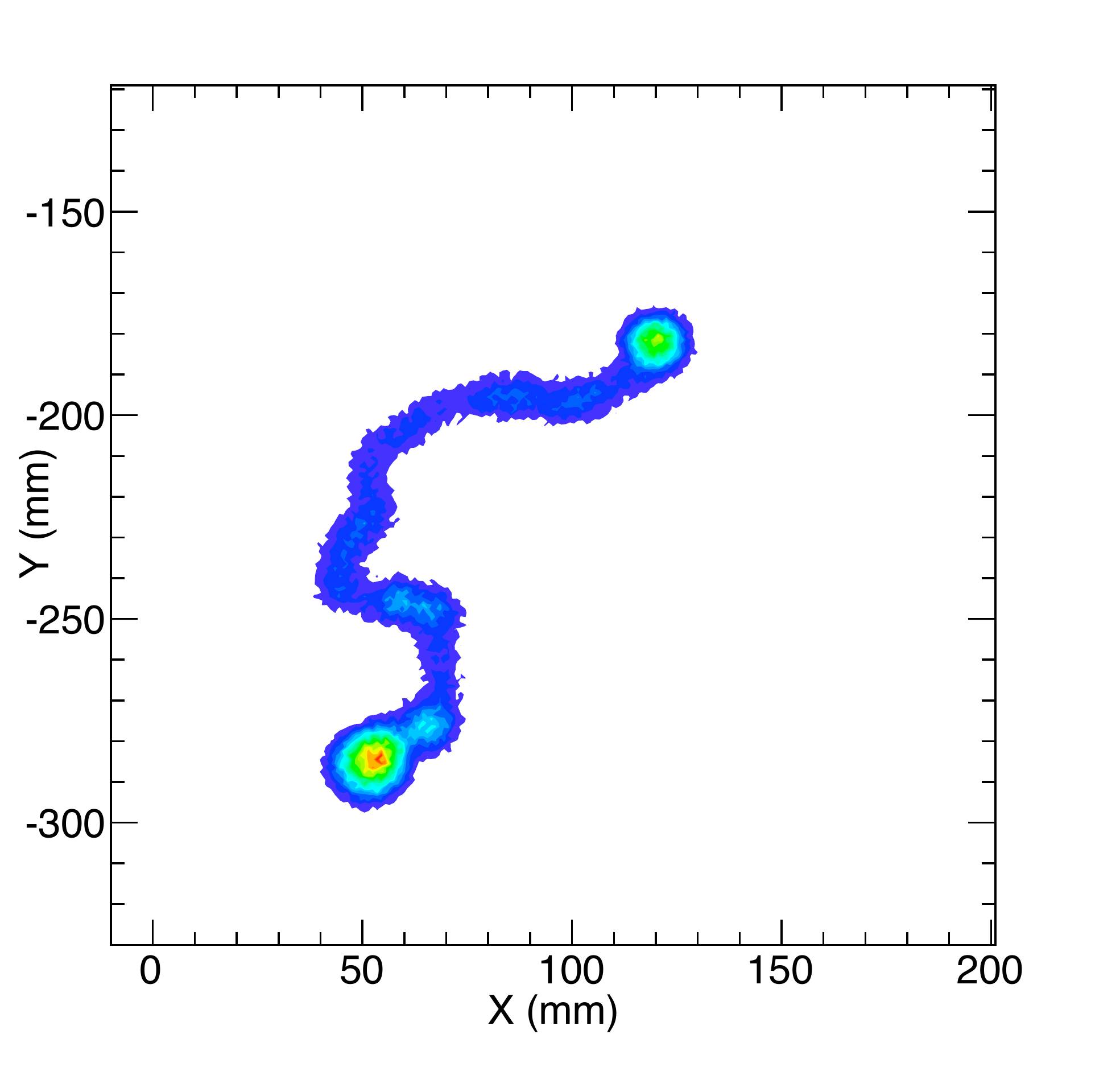}
\caption{Monte-Carlo simulation of a \XE\ \bbonu\ event in xenon gas at 10 bar: the ionization track, about 30 cm long, is tortuous because of multiple scattering, and has larger depositions or \emph{blobs} in both ends.}
\label{fig:track}
\end{figure}

Neutrinoless double beta decay events leave a distinctive topological signature in gaseous xenon: an ionization track, about 30 cm long at 10 bar, tortuous due to multiple scattering, and with larger energy depositions at both ends (see figure \ref{fig:track}). The Gotthard experiment \cite{Luscher:1998sd}, consisting in a small xenon gas TPC (5.3 kg enriched to 68\% in \XE) operated at 5 bar, proved the effectiveness of such a signature to discriminate signal from background.

%%%%%%%%%%
%%%%%%%%%%

\begin{figure}
\centering
\includegraphics[width=0.7\textwidth]{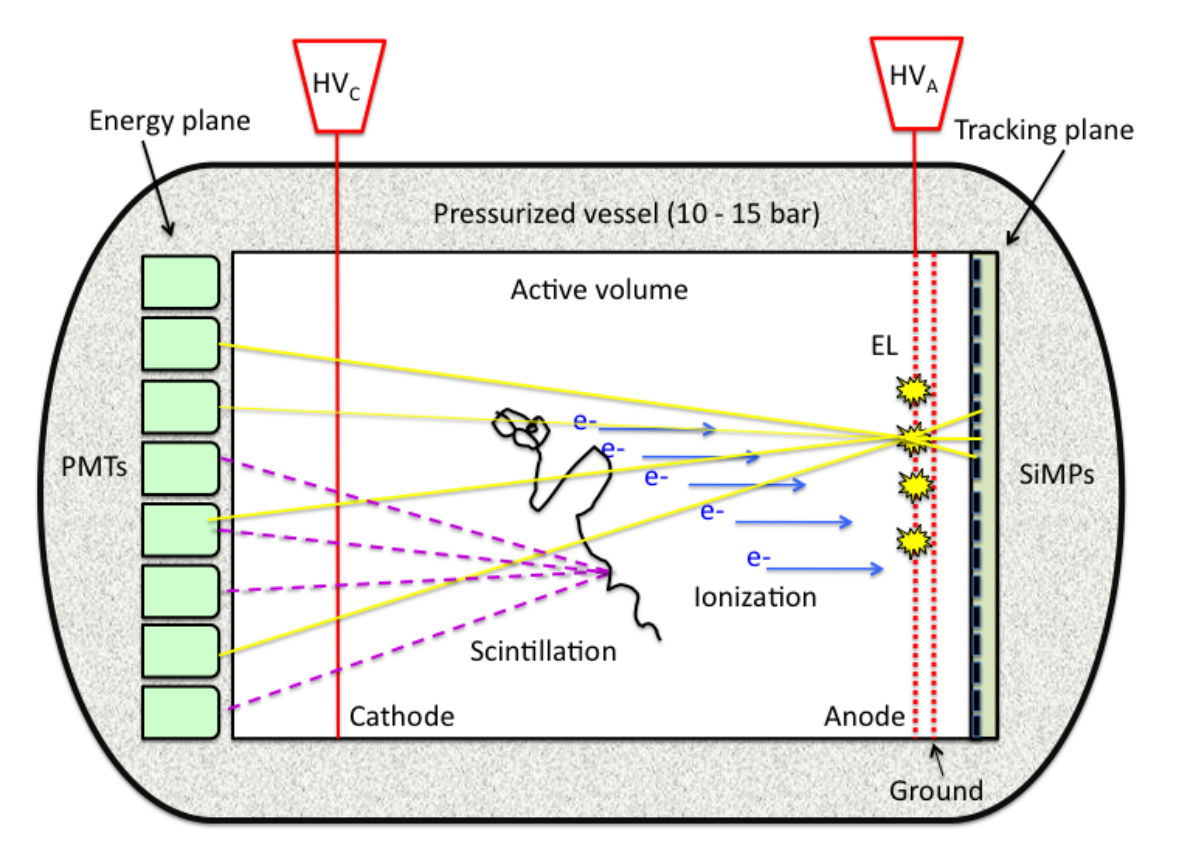}
\caption{The \emph{Separate, Optimized Functions} (SOFT) concept in the NEXT experiment: EL light generated at the anode is recorded in the photosensor plane right behind it and used for tracking; it is also recorded in the photosensor plane behind the transparent cathode and used for a precise energy measurement.} \label{fig:SOFT}
\end{figure}

To achieve optimal energy resolution, the ionization signal is amplified in NEXT using the electroluminescence (EL) of xenon. Also, following ideas introduced in \cite{Nygren:2009zz} and further developed in our \emph{Letter of Intent} \cite{Granena:2009it} and CDR \cite{Alvarez:2011my}, the chamber will have separated detection systems for tracking and calorimetry. This is the so-called \emph{SOFT} concept, illustrated in figure \ref{fig:SOFT}. The detection process is as follows: Particles interacting in the HPXe transfer their energy to the medium through ionization and excitation. The excitation energy is manifested in the prompt emission of VUV ($\sim178$ nm) scintillation light. The ionization tracks (positive ions and free electrons) left behind by the particle are prevented from recombination by an electric field (0.3--0.5 ${\rm kV}/{\rm cm}$). The ionization electrons drift toward the TPC anode, entering a region, defined by two highly-transparent meshes, with an even more intense electric field (3 ${\rm kV}/{\rm cm}/{\rm bar}$). There, further VUV photons are generated isotropically by electroluminescence. Therefore both scintillation and ionization produce an optical signal, to be detected with a sparse plane of PMTs (the \emph{energy plane}) located behind the cathode. The detection of the primary scintillation light constitutes the start-of-event, whereas the detection of EL light provides an energy measurement. Electroluminescent light provides tracking as well, since it is detected also a few millimeters away from production at the anode plane, via an array, 1 cm pitch, of 1-mm$^{2}$ MPPCs (the \emph{tracking plane}).

%%%%%%%%%%
%%%%%%%%%%

%%%%%%%%%%%%%%%%%%%%%%%%%%%%%%%%%%%%%%%%%%%%%%%%%%%%%%%%%%%%
\subsection{Development of the NEXT project: R\&D and prototypes}
\begin{figure}
\centering
\includegraphics[width=0.475\textwidth]{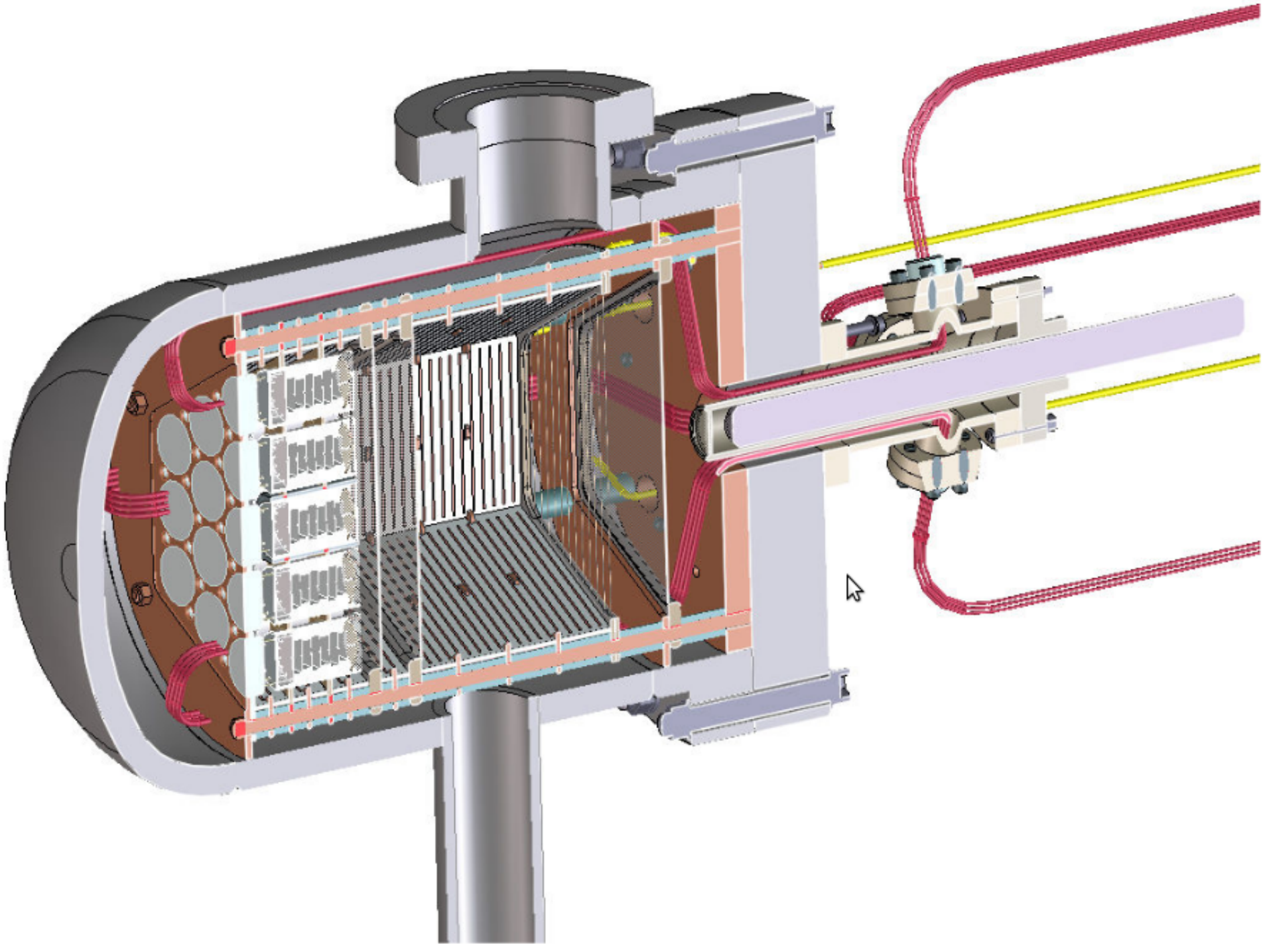}
\includegraphics[width=0.475\textwidth]{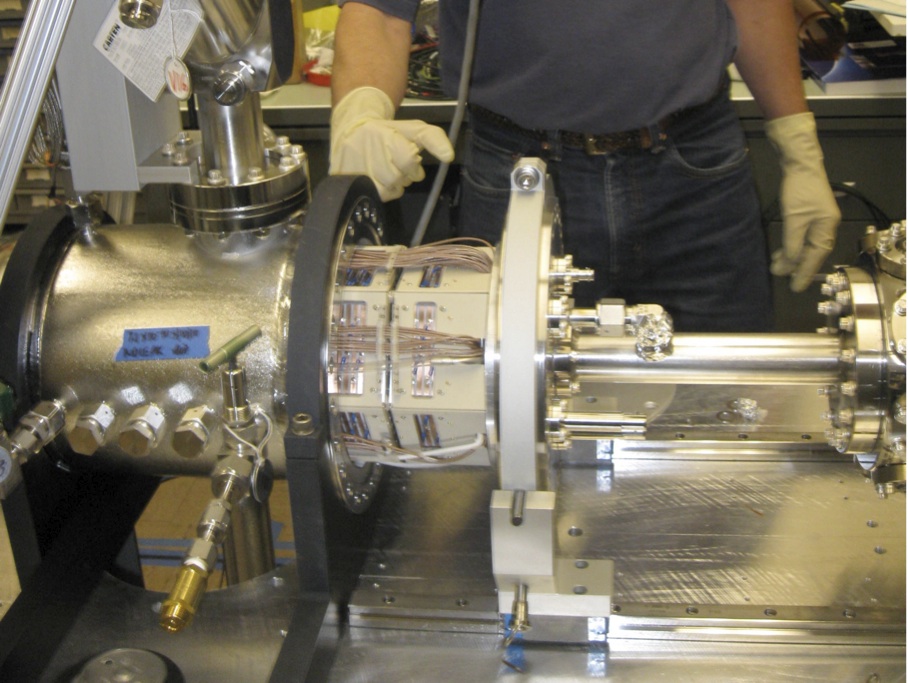}
\caption{The NEXT-DBDM prototype, operating at LBNL. Left: cross section of the detector. Right: insertion of the time projection chamber into the stainless-steel pressure vessel.} \label{fig:DBDM}
\end{figure}

\begin{figure}
\centering
\includegraphics[height=8.cm]{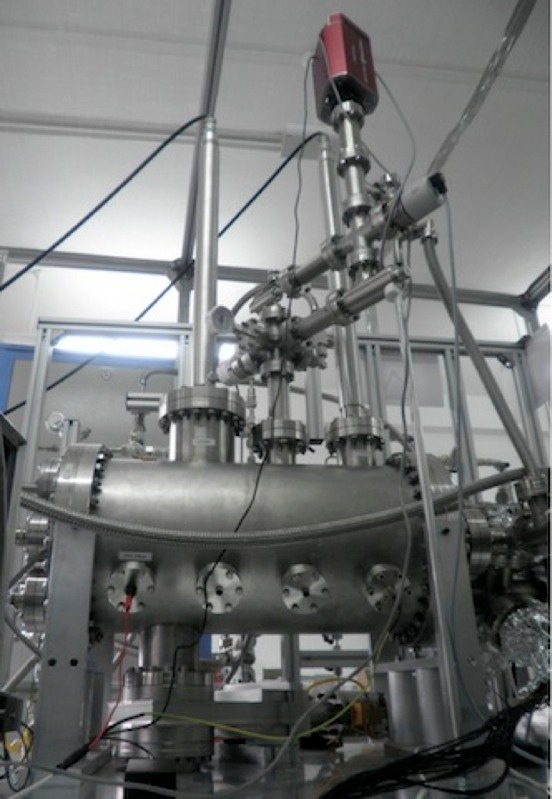}
\includegraphics[height=8.cm]{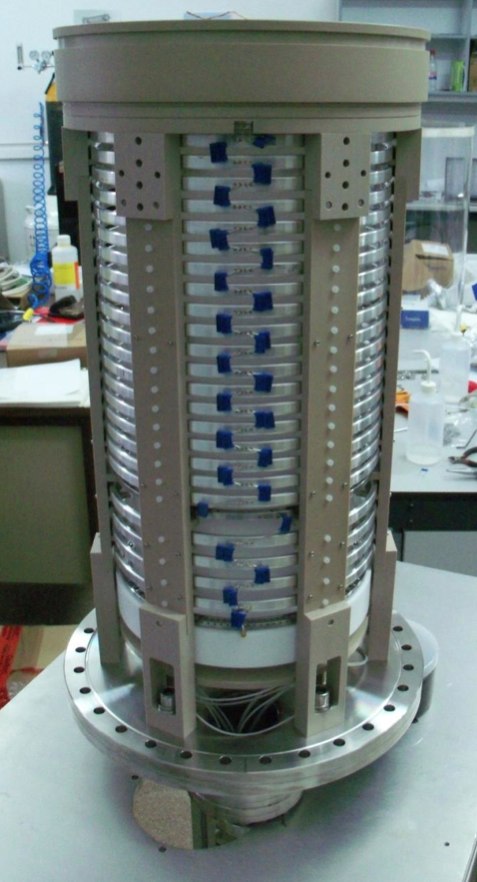}
\caption{The NEXT-DEMO prototype, operating at IFIC. Left: the pressure vessel. Right: the time projection chamber.} \label{fig:DEMO}
\end{figure}

During the last three years, the NEXT R\&D program has focused in the construction, commissioning and operation of three prototypes:
\begin{itemize}
\item \emph{NEXT-DBDM}, shown in figure \ref{fig:DBDM}. This is an electroluminescent TPC equipped with a compact array of 19 Hamamatsu R7378A PMTs (1-inch size, sensitive to VUV light and operable at pressures up to 17 bar). The fiducial volume, a hexagonal prism, is 8 cm long with 17 cm diameter. The detector vessel, a 10 L stainless-steel container, can hold about 1 kg of xenon gas at 15 bar. The main goal of this prototype was to perform detailed energy resolution studies. The detector is operating at LBNL.

\item \emph{NEXT-DEMO}, shown in figure \ref{fig:DEMO}. This larger prototype, operating at IFIC, is equipped with an energy plane made of 19 Hamamatsu R7378A and a tracking plane made of $\sim300$ 1-mm$^{2}$, Hamamatsu MPPCs. The stainless-steel pressure vessel, 60 cm long and 30 cm diameter, can withstand 15 bar. The main goal of the prototype was the demonstration of the detector concept to be used in NEXT-100. More specifically: (a) to demonstrate track reconstruction and the performance of MPPCs; (b) to test long drift lengths and high voltages (up to 50 kV in the cathode and 25 kV in the anode); (c) to understand gas recirculation in a large volume, including operation stability and robustness against leaks; (d) to understand the transmittance of the light tube, with and without wavelength shifter. 

\item \emph{NEXT-MM}, a prototype initially used to test the Micromegas technology and currently used to explore new gas mixtures. NEXT-MM operates at the University of Zaragoza.
\end{itemize}

\begin{figure}
\centering
\includegraphics[width=0.495\textwidth]{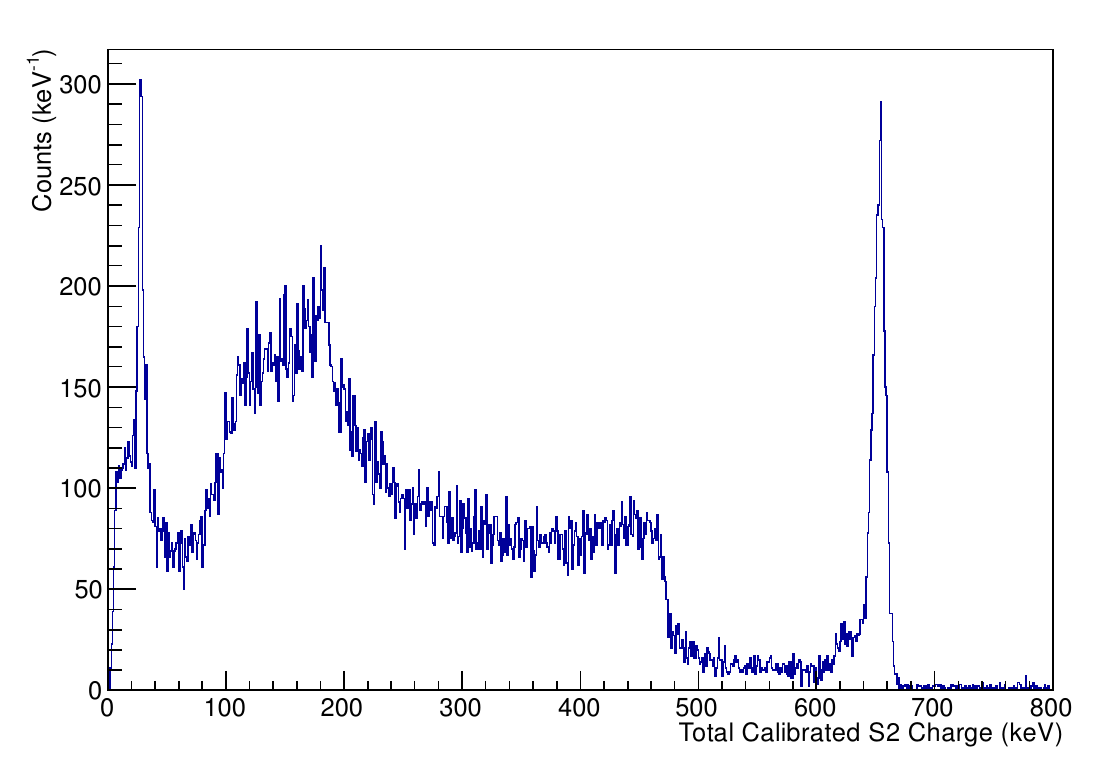} 
\includegraphics[width=0.495\textwidth]{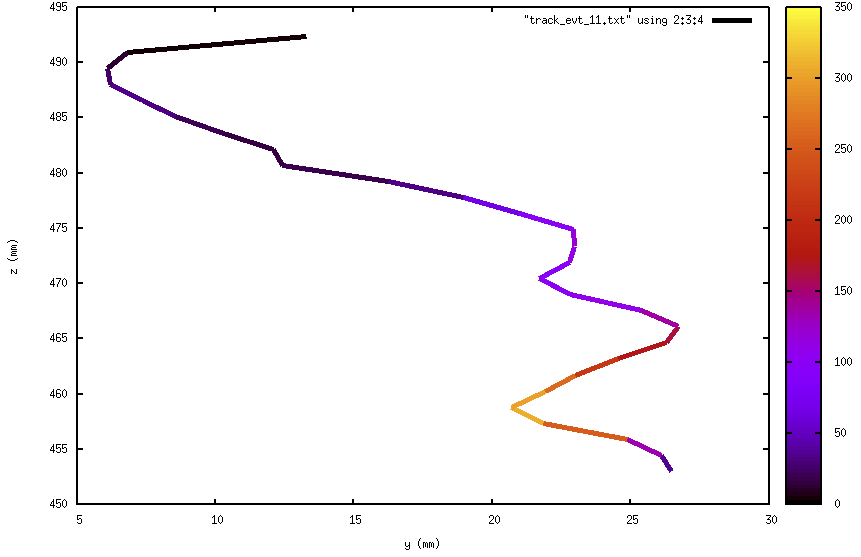} 
\caption{Left: Energy spectrum of \CS\ measured by NEXT-DBDM using a calibration source. The photoelectric peak at 662 keV is measured with a resolution close to 1\% FWHM \cite{Renner:2011}. This extrapolates to $\sim$0.5\% FWHM at the energy of the \Qbb\ of \XE. Right: The reconstructed track left by a photoelectric electron produced by the interaction of a 662-keV gamma (from a \CS\ calibration source) detected by NEXT-DEMO. Abscises and ordinates represent the transverse coordinates of the chamber; the color scale indicates energy. The line connects points (central values) successively reconstructed in time; the associated uncertainty (of the order of 5--10 millimeters) is not shown here.} \label{fig:results} 
\end{figure}

The initial results of the prototypes show already excellent energy resolution and tracking capabilities, see figure \ref{fig:results}. Several papers are in preparation and will be published during 2012.

%%%%%%%%%%
%%%%%%%%%%

%%%%%%%%%%
%%%%%%%%%%

%%%%%%%%%%
%%%%%%%%%%

%%%%%%%%%%%%%%%%%%%%%%%%%%%%%%%%%%%%%%%%%%%%%%%%%%
\subsection{The NEXT-100 detector}
\begin{figure}
\centering
\includegraphics[width=0.9\textwidth]{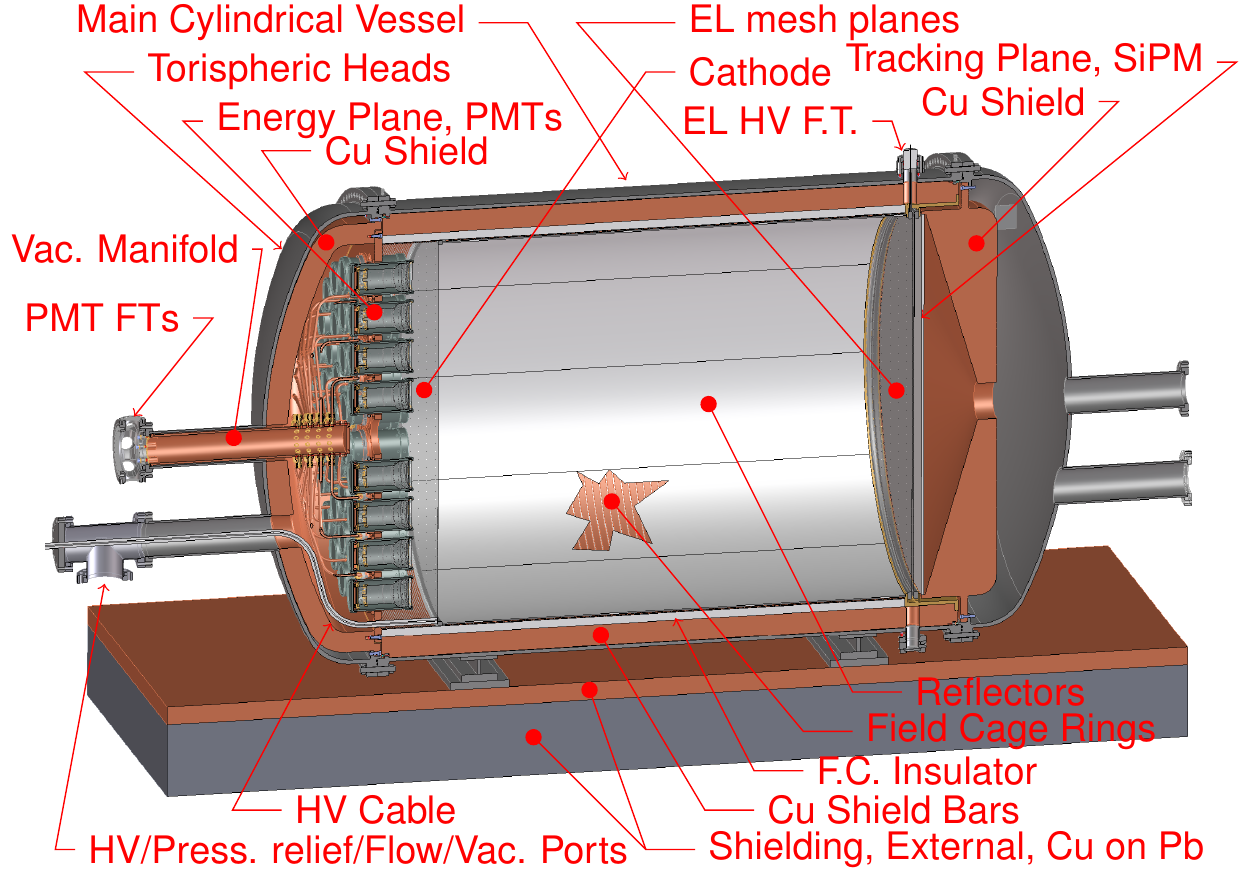} 
\caption{The NEXT-100 detector.} \label{fig:NEXT100} 
\end{figure}

Figure \ref{fig:NEXT100} shows a sketch of the NEXT-100 detector, indicating all the major subsystems. These are: 
\begin{itemize}
\item The pressure vessel (PV) --- described in section \ref{sec:PressureVessel} ---, built in stainless steel and able to hold 15 bar of xenon. A copper layer on the inside shields the sensitive volume from the radiation originated in the vessel material. 
\item The field cage (FC), electrode grids, HV penetrators and light tube, described in section \ref{sec:FieldCage}.
\item The tracking plane (TP) made of MPPCs arranged into dice boards (DB). The front-end electronics is inside the gas, shielded behind a thick copper plate (section \ref{sec:TrackingPlane}).
\item The energy plane (EP) made of PMTs housed in copper enclosures (section \ref{sec:EnergyPlane}).
\item The lead castle that shields the detector against external $\gamma$-rays, described in section \ref{sec:Shielding}.
\end{itemize}

%%%%%%%%%%
%%%%%%%%%%

In addition, the electronics and data-acquisition system are outlined in section \ref{sec:Electronics}; the xenon gas system is described in section \ref{sec:GasSystem}; section \ref{sec:Infrastructure} deals with the installation of the detector at the LSC; and the background model of NEXT-100 is discussed in section \ref{sec:Background}.

\section{The pressure vessel} \label{sec:PressureVessel}
\begin{figure}
\centering
\includegraphics[width=\textwidth]{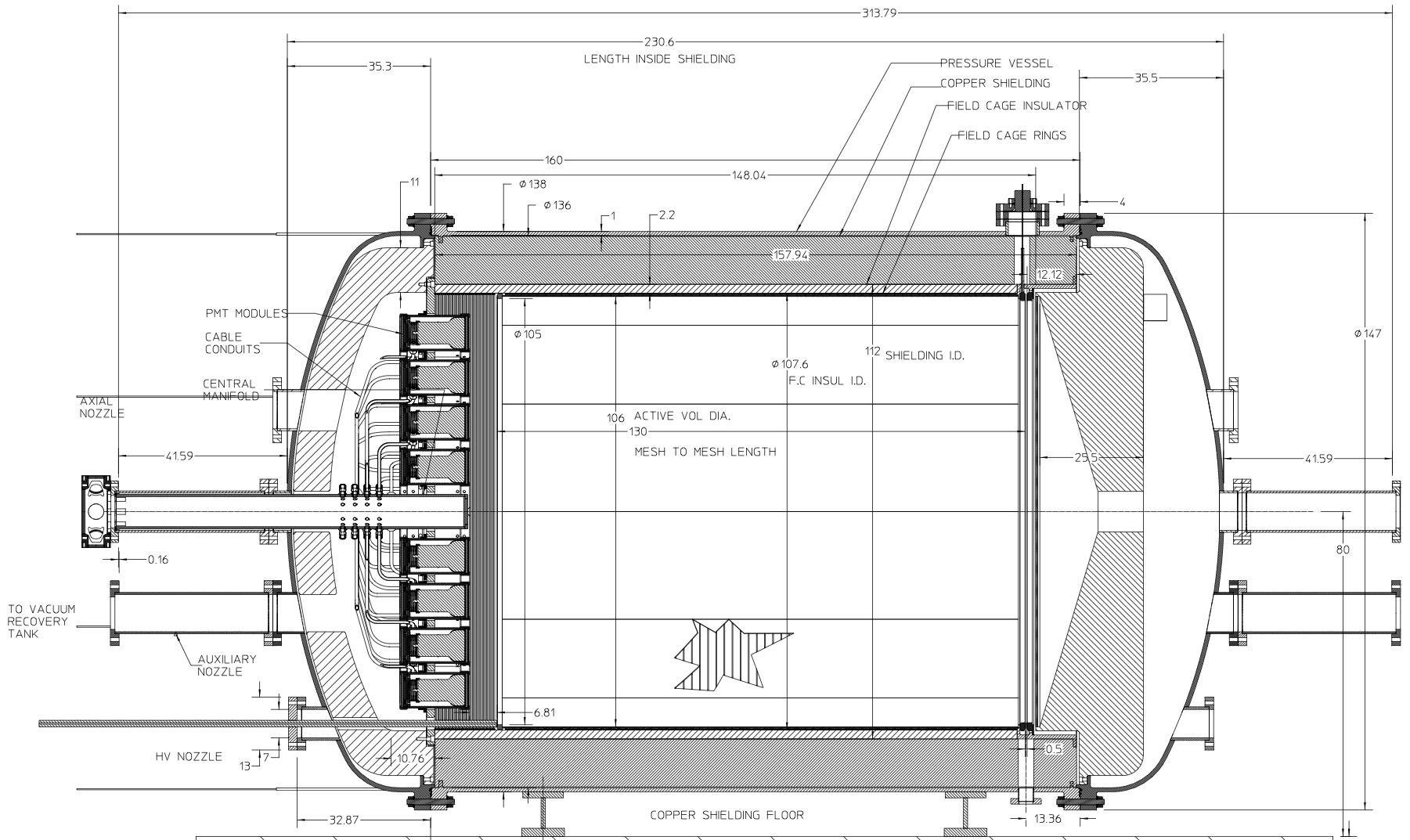}
\caption{Pressure Vessel/Detector: side cross section view.} \label{fig:PV} 
\end{figure}

The pressure vessel (PV) consists of a cylindrical center section (barrel) with two identical torispheric heads on each end, their main flanges bolted together.  The vessel orientation is  horizontal, so as to minimize the overall height; this reduces the outer shielding cost and allows essentially unlimited length on each end for cabling and service expansion. Table \ref{tab:PV} collects the basic parameters and dimensions of the PV, and a longitudinal cross section is shown in figure \ref{fig:PV}.

%%%%%%%%%%
\begin{table}
\caption{NEXT-100 pressure vessel basic parameters and dimensions (some of these quantities could change slightly during the construction phase due to refinements in the design).} \label{tab:PV}
\begin{center}
\begin{tabular}{ll}
\toprule
Maximum operating pressure (differential) & 14.0 bar \\
Maximum allowable working pressure (differential) & 15.4 bar \\ 
Maximum allowable external pressure (differential) & 1.5 bar \\ \midrule
Inner diameter & 136 cm \\
Outer diameter, barrel and heads & 138 cm \\
Outer diameter, main flanges & 147 cm \\
Length, head to head, inside & 228 cm \\
Thickness, barrel and head wall & 10 mm \\
Thickness, main flanges (each side) & 4.0 cm \\ \midrule
Number of bolts, main flanges, each set & 140 \\
Bolt diameter, main flanges & 14 mm \\
Bolt length, main flanges & 11 cm \\ \midrule
Mass, pressure vessel & 1\,200 kg \\
Mass, internal copper shielding (including heads) & 10\,000 kg \\
Mass, energy plane & 750 kg \\
Mass, field cage & 250 kg\\
Mass, tracking plane & 300 kg \\
Mass, NEXT-100 total & 12\,500 kg \\
\bottomrule
\end{tabular}
\end{center}
\end{table}
%%%%%%%%%%

%%%%%%%%%%
%%%%%%%%%%

The vessel will be made of stainless steel, specifically the low-activity 316Ti alloy, unless similarly low activity 304L or 316L alloys can be found which will allow use of roll forgings for the flanges (these promise better leak tightness and mechanical integrity). Measurements by the XENON collaboration show that it is possible to secure 316Ti with an activity at the level of 0.2 ${\rm mBq}/{\rm kg}$ for the thorium series and 1.3 ${\rm mBq}/{\rm kg}$ for the uranium series \cite{Aprile:2011ru}. The mass of the PV is 1\,200 kg, resulting in a total activity of about 1.6 Bq for the uranium series. To shield this activity we introduce an \emph{inner copper shield} (ICS) 12 cm thick and made of radiopure copper, with an activity of about 5--10 $\mu$Bq/kg. The ICS will attenuate the radiation coming from the external detector (including the PV and the external lead shield) by a factor of 100. 
After the ICS the residual activity due to the PV is about 0.02 Bq. One needs to add the residual activity of the ICS itself which is, taking into account self-shielding, of the order of 0.03 Bq. Thus, the resulting activity of the whole system is $\sim0.05$ Bq.  

The PV supports a number of internal components on the main flanges; nothing is supported directly on the 10 mm thick shell. The barrel flanges have an inside flange containing a circle of 240 M8 threaded holes. The internal copper shield bars attach on each end to these internal flanges; in turn these bars have machined features which then support the field cage and the sensor planes. The heads also have an internal flange; to these are fastened the internal copper shield plates. The cathode plane high voltage feedthrough is integrated into the energy head and makes contact with the cathode plane when the head is assembled. 

Both heads have four axial nozzles, each with a 100 mm outer diameter and a 150 mm diameter flange. The central nozzle of each head is for services (power and signal cabling) to the sensor planes. The two auxiliary nozzles on each side of this are for gas flow and pressure relief; at present only one on each head is used. The fourth nozzle, located furthest from the vessel axis, is used for the high voltage feedthrough of the cathode plane. To keep heads identical, one will be present, but capped off on the tracking head, to allow for future reconfiguration. All these axial nozzles are located on the vertical midplane of the vessel; this allows the two lead shielding walls to come together at this midplane by making semicircular cutouts on their mating surfaces. There is also a ring of eight \emph{radial} nozzles, located at the EL gap. Two of these (one on top, and one at a 45 deg angle) are for the EL gate high voltage feedthrough (2 possible locations) and the other 6 are inspection ports for the EL gap; they may prove useful for EL diagnostics and in-situ cleaning or repair. All nozzles will be as short as possible, with simple flat-faced flanges. This is to reduce vessel damage risk, to ease fabrication, and to preserve flexibility in length; extension spools are used to bring services out through the shielding. Nozzle flanges, although being flat-faced (with the double O-rings on the spools and caps) use standard CF flange bolt hole pattern; this is to preserve the possibility of using CF components (pressure rated). These could be pre-assembled to  double O-ring/CF adapter plates, then pressure and leak tested  prior to assembling on the nozzle.

All pressure sealing flange joints that are exposed to atmosphere on the outside are sealed using double O-rings in grooves, for both sealing reliability, and to minimize the flange and bolt sizes. The inner O-ring is for pressure sealing; the outer O-ring serves not only as a backup, but also to create a sealed annulus which can be continuously monitored for leakage by pulling a vacuum on it with an RGA monitor (sense port). Xenon will permeate through these O-rings and will need to be recovered in a cold trap. The total amount is estimated to be $<200\ {\rm gram}/{\rm year}$  for butyl or nitrile O-rings; this includes the nozzle flanges and PMT enclosure O-ring leakage. The use of metal C-ring gaskets (special low force design) is being explored. The standard sealing force for these gaskets is very high and greatly increases flange thickness and diameter. The pressure vessel flanges are being designed with some reserve capacity for the higher bolt force that would be required for these low force metal gaskets.

The head to vessel flange bolts are Inconel 718; this is the highest strength noncorrosive bolting material allowed in the ASME code. Flange thickness and outer diameter are substantially minimized by using the highest possible strength bolting; this mostly compensates for the lower radiopurity of this material (relative to Inconel 625, the next best alloy) and saves on external shielding cost.  

Saddle supports will be welded to the barrel wall, as is standard practice. The barrel will be bolted through the shielding floor to the seismic platform below. The head/shield assemblies will first be attached to a precisely adjustable 6-strut support fixture, also known as a \emph{hexapod} or \emph{Stewart platform}. This is a cradle support ring for the head which is connected to a baseplate by adjustable length turnbuckle struts, forming a kinematic mount.  This fixture is attached to a set of linear ball bearing carriages that slide on precision rails that are bolted to the shielding floor when needed. There is 1 m carriage travel to allow each head to be retracted far enough to clear the sensor planes inside. The 6-strut fixture is easily adjusted by hand to mate the head flanges to the barrel flanges with very high precision, whereby they can be bolted together without stress. The adjustments are essentially independent for the three degrees of translation and for roll and yaw about the vessel axis. The pitch adjustment is coupled only with the Z translation (along the axis) which is easy to deal with since this is also the rail motion direction. This 6-strut fixture will also be used to assemble the PMT and SiPM arrays to their attachment points in the barrel.

The vessel will be built strictly to \emph{ASME Pressure Vessel Design Code, Section VIII}. It has been designed almost entirely by the Collaboration; the outside fabricator will be required however to supply additional details of fabrication, such as weld design, for approval, as these can be dependent on the in-house capabilities. The fabricator is also responsible for pressure integrity, under ASME rules, and will likely perform their own calculations to verify soundness of design. The Collaboration will be specifying many aspects (and approving every aspect) of the design and fabrication, such as: cleaning of joints, weld preparation methods, inspection methods, fabrication sequence, post-weld heat treatments, etc. This is to assure that the unusually high tolerances on dimensions and radiopurity are met, without needing rework.

\section{The field cage} \label{sec:FieldCage}
The main body of the field cage will be a high-density polyethylene (HDPE) cylindrical shell, 2.5 cm thick, that will provide electric insulation from the vessel. Three wire meshes --- cathode, gate and anode --- separate the two electric electric field regions of the detector (see table \ref{tab:FC}). The drift region, between cathode and gate, is a cylinder of 107 cm diameter and 130 cm length. Copper strips attached to the HDPE and connected with low background resistors grade the high voltage. The EL region, between gate and anode, is 0.5 cm long. There is also a buffer region between the cathode and the energy plane which will be used to degrade the high voltage in the cathode safely to ground.

%%%%%%%%%%
\begin{table}
\caption{Basic parameters of the electric field regions (drift and EL) of NEXT-100.} \label{tab:FC}
\begin{center}
\begin{tabular}{ll}
\toprule
Drift field strength & 0.3 kV cm$^{-1}$ \\
EL field strength ($E/p$) & 3.0 kV cm$^{-1}$ bar$^{-1}$ \\
Optical gain & 2500 ${\rm photons}/\mathrm{e}^{-}$ \\
Drift length & 130 cm \\
EL gap & 0.5 cm \\
Cathode voltage & $-58$ kV \\
Gate grid voltage & $-22.5$ kV \\
Anode grid voltage & 0 \\
\bottomrule
\end{tabular}
\end{center}
\end{table}
%%%%%%%%%%

\begin{figure}
\centering
\includegraphics[height=7cm]{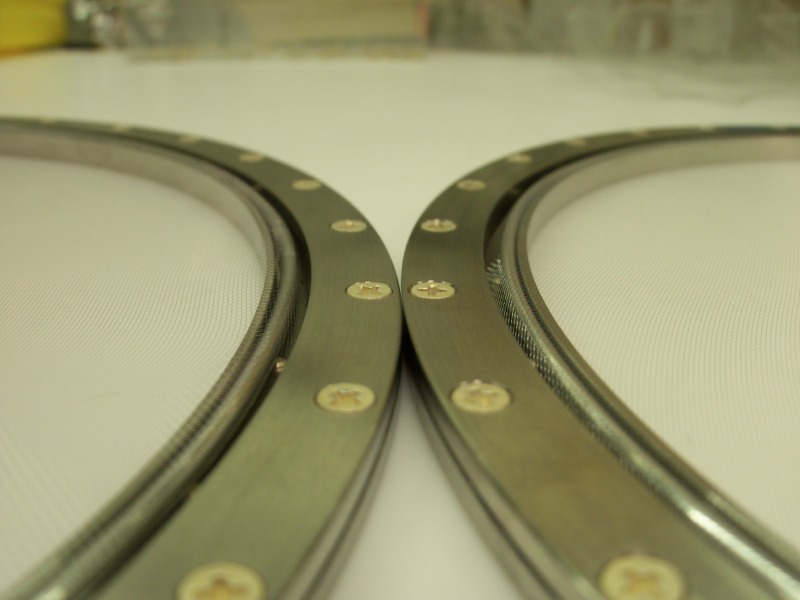}
\caption{Detail of the wire meshes used in NEXT-DEMO.} \label{fig:grids}
\end{figure}

Cathode, anode and gate will be similar to the wire meshes used in NEXT-DEMO, shown in figure \ref{fig:grids}. They were constructed using stainless steel mesh with 30-$\mu$m wire diameter and a 0.5 mm wire pitch, which results in an open area of 88\%. The grids are formed by clamping in a ring with a tongue and groove to hold the mesh and using a tensioning ring that is torqued with set screws to achieve the optimum tension. For the large diameter required in NEXT-100, preliminary estimates show that electrostatic attraction will cause the EL grids to bow considerably; this can be remedied by using a larger gauge wire. For example, wire mesh with a similar open area is available with 90 micron wire diameter made from titanium. We are also investigating the use of titanium or copper grid frames to minimize the radioactive budget. However, the total mass of the EL and HVFT system is small and the use of low-background steel will be sufficient to keep the radioactive budget acceptable. 

%%%%%%%%%%
%%%%%%%%%%

\begin{figure}
\centering
\includegraphics[width=0.8\textwidth]{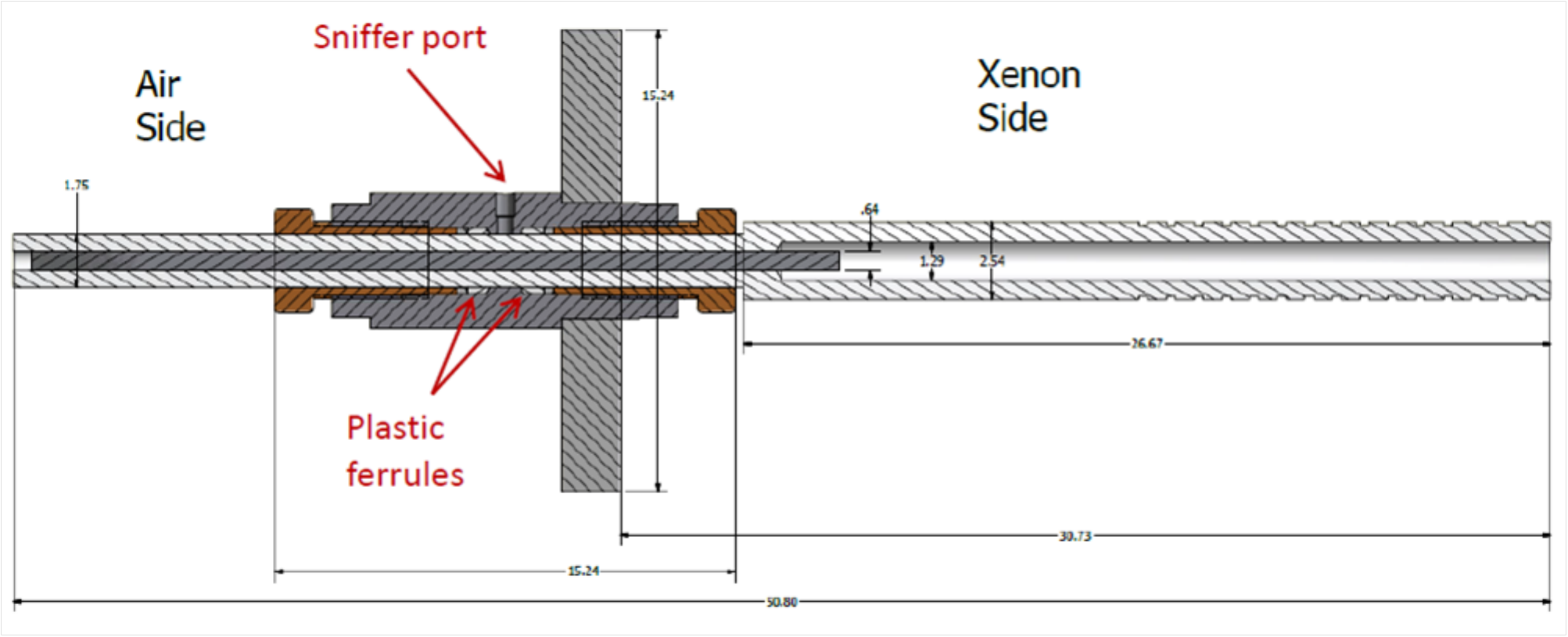}
\caption{Cathode high-voltage feedthrough (HVFT) designed for operation up to 100 kV.} \label{fig:feedthru}
\end{figure}

The cathode high voltage feedthrough (HVFT) will be constructed using a compression seal approach, as illustrated in figure \ref{fig:feedthru}.  A metal rod is pressed into a plastic tube (Tefzel or FEP, which have high dielectric strength) which is then clamped using plastic ferrules from both the pressure side and air side.  A sniffer port is placed between the seals to assure that xenon is not leaking.  The feedthrough will be attached to a flange located on the energy plane end-cap.  A shielded cable will be connected to the feedthrough and placed through the PMT support plate.  The unshielded portion of the cable, with an additional resistive coating, will then run along the inside of the buffer field rings and mate with the cathode via a spring loaded junction.  This approach, with the exception of the resistive coating, has been used in NEXT-DEMO, where a cathode voltage of 45 kV has been achieved.  A smaller prototype was tested to 100 kV in vacuum and 70 kV in nitrogen at 3 bar. It has been demonstrated to be xenon leak tight at 10 bar and 10\textsuperscript{-7} mbar vacuum.  

%%%%%%%%%%
%%%%%%%%%%

\begin{figure}
\centering
\includegraphics[width=0.7\textwidth]{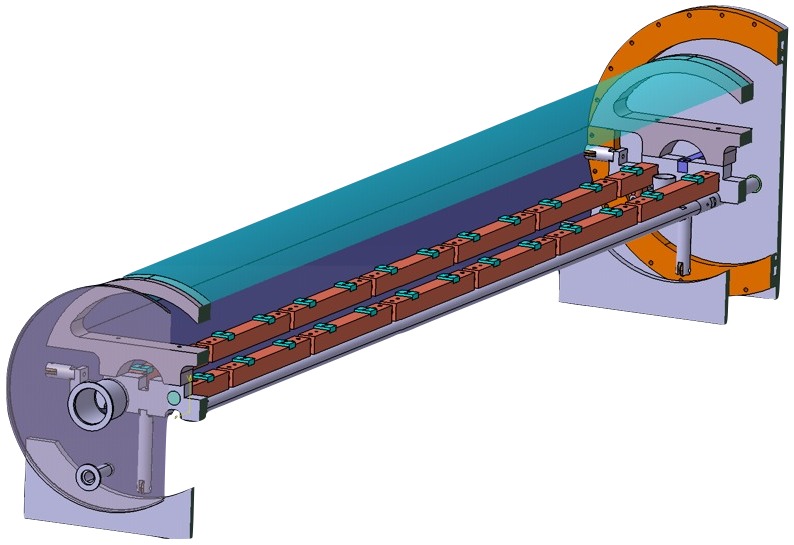} \\[2pt]
\includegraphics[height=7cm]{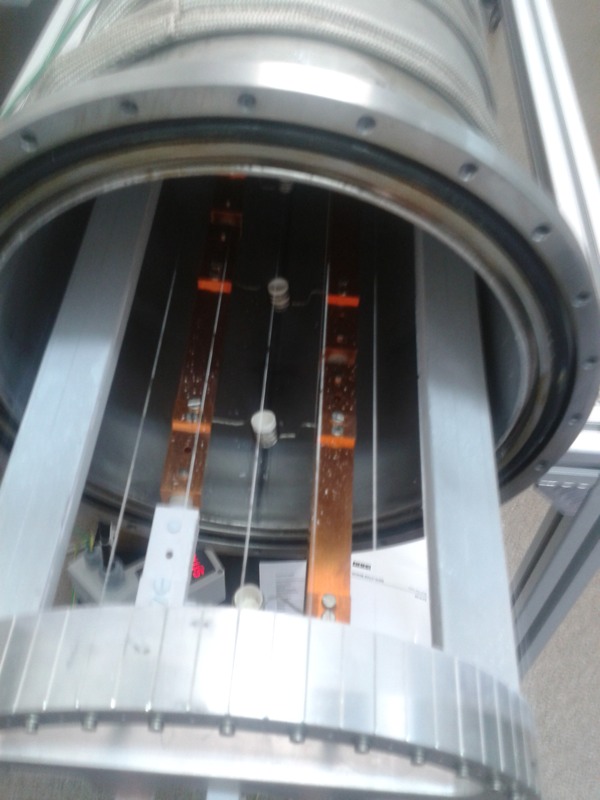}
\includegraphics[height=7cm]{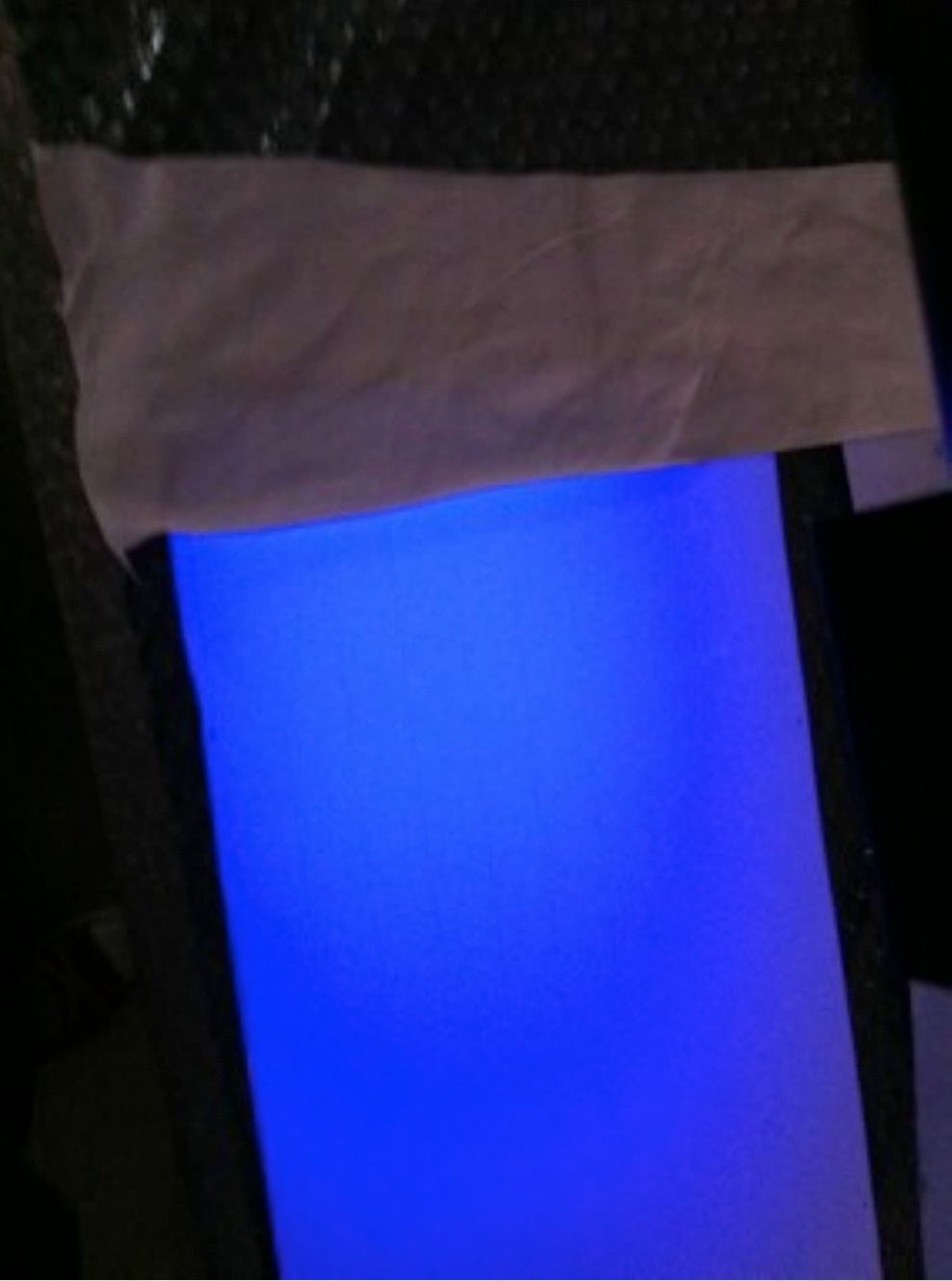}
\caption{Top: Sketch of the evaporator used in the preparation of the field cage reflectors. Bottom left: Detail of the evaporator crucibles, where the TPB powder is deposited. The cables used to hold the samples are also visible. Bottom right: Response of a TTX sheet to VUV light after coating, showing the characteristic blue-light emission when illuminated by a UV lamp. In the upper part of the picture, the same foil without coating can be seen.}
\label{fig:wlsfoil}
\end{figure}

To improve the light collection efficiency of the detector, reflector panels coated with a wavelength shifter will cover the inner part of the field cage. We have chosen the solution developed by the ArDM collaboration \cite{Boccone:2009kk}: sheets of Tetratex$^{\textregistered}$ (TTX) --- an aligned polytetrafluoroethylene (PTFE) fibrous cloth with nearly a 100\% diffuse Lambertian reflectivity --- are vacuum-coated with tetraphenyl butadiene (TPB), a standard wavelength shifter from VUV to blue, resulting in a reflectivity of about 97\% in the blue region of the spectrum. Figure \ref{fig:wlsfoil} (bottom right) shows a coated TTX sheet inspected optically with a UV lamp, with the characteristic re-emission in the blue. The NEXT collaboration has acquired the large evaporation chamber developed by the ArDM collaboration \cite{Boccone:2009kk}, and it is currently installed at IFIC. This is a stainless steel vacuum chamber (see figure \ref{fig:wlsfoil}) large enough to house TTX sheets of 120$\times$25 cm$^2$, which have the right size for the NEXT-100 detector. 

%%%%%%%%%%
%%%%%%%%%%

\section{The tracking plane} \label{sec:TrackingPlane}
The tracking function in NEXT-100 will be provided by a plane of multi-pixel photon counters (MPPCs) operating as sensor pixels and located behind the transparent EL gap. The chosen MPPC is the S10362-11-050P model by Hamamatsu \cite{MPPC}. This device has an active area of 1 mm$^{2}$, 400 sensitive cells (50 $\mu{\rm m}$ size) and high photon detection efficiency in the blue region (about $\sim50\%$ at 440 nm). The dark count rate is 0.4 MHz, that is, less than 1 event per microsecond (which is the considered sampling time). This random noise events have amplitudes of up to 8 photoelectrons, and thus a digital threshold at those levels should lead to an insignificant noise rate in NEXT-100 without affecting the tracking performance. We have measured the spread in gain between the sensors to be less than 4\%; this ensures a homogenous response of the plane in the reconstruction of tracks. Last but not least, MPPCs are very cost-effective and their radioactivity is very low, given its composition (mostly silicon) and very small mass.  

\begin{figure}
\centering
\includegraphics[width=0.7\textwidth]{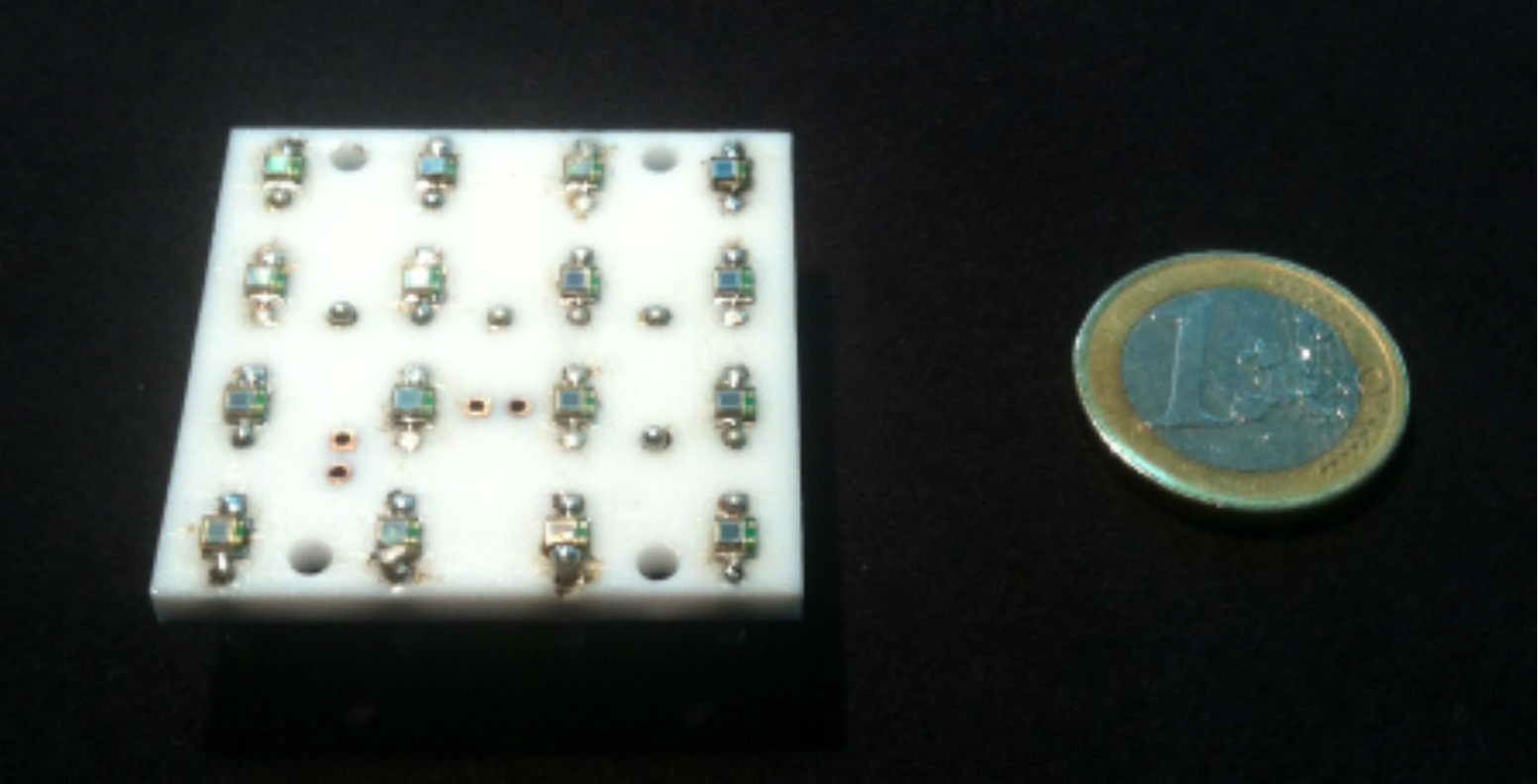}
\caption{Dice Board, used in NEXT-DEMO, containing 16 (4$\times$4) MPPCs.} \label{fig.db2}
\end{figure}

The MPPCs will be mounted in Dice Boards (DB). These are square boards made of {\em cuflon} (PTFE fixed to a copper back plane). Figure \ref{fig.db2} shows one of the DB of the NEXT-DEMO prototype, holding 4$\times$4 MPPCs. NEXT-100 DBs will be similar but containing 8$\times$8 sensors. 

%%%%%%%%%%
%%%%%%%%%%

The pitch of the NEXT-100 tracking plane is a compromise between several constraints imposed by physics. Charge diffusion in pure xenon gas is large; for electric field strengths around 0.5 kV/cm, the transversal \emph{diffusion coefficient} is about 1 ${\rm mm}/\sqrt{\rm cm}$. Therefore a pitch significantly smaller than 1\,cm is not useful. Conversely, as the pitch increases, the background rejection capabilities of the tracking function decrease due to the worse spatial resolution and two-track separation. Simulations show that a reasonable tradeoff may be found for pitches in the range 1--1.5 cm. While physics performance appears not to degrade too much with pitch in that region, the number of pixels decreases with the square of the pitch. A reasonable compromise appears to be 1.1\,cm, which in turn requires about 7\,000 pixels.

\begin{figure}
\centering
\includegraphics[width= 6.5cm]{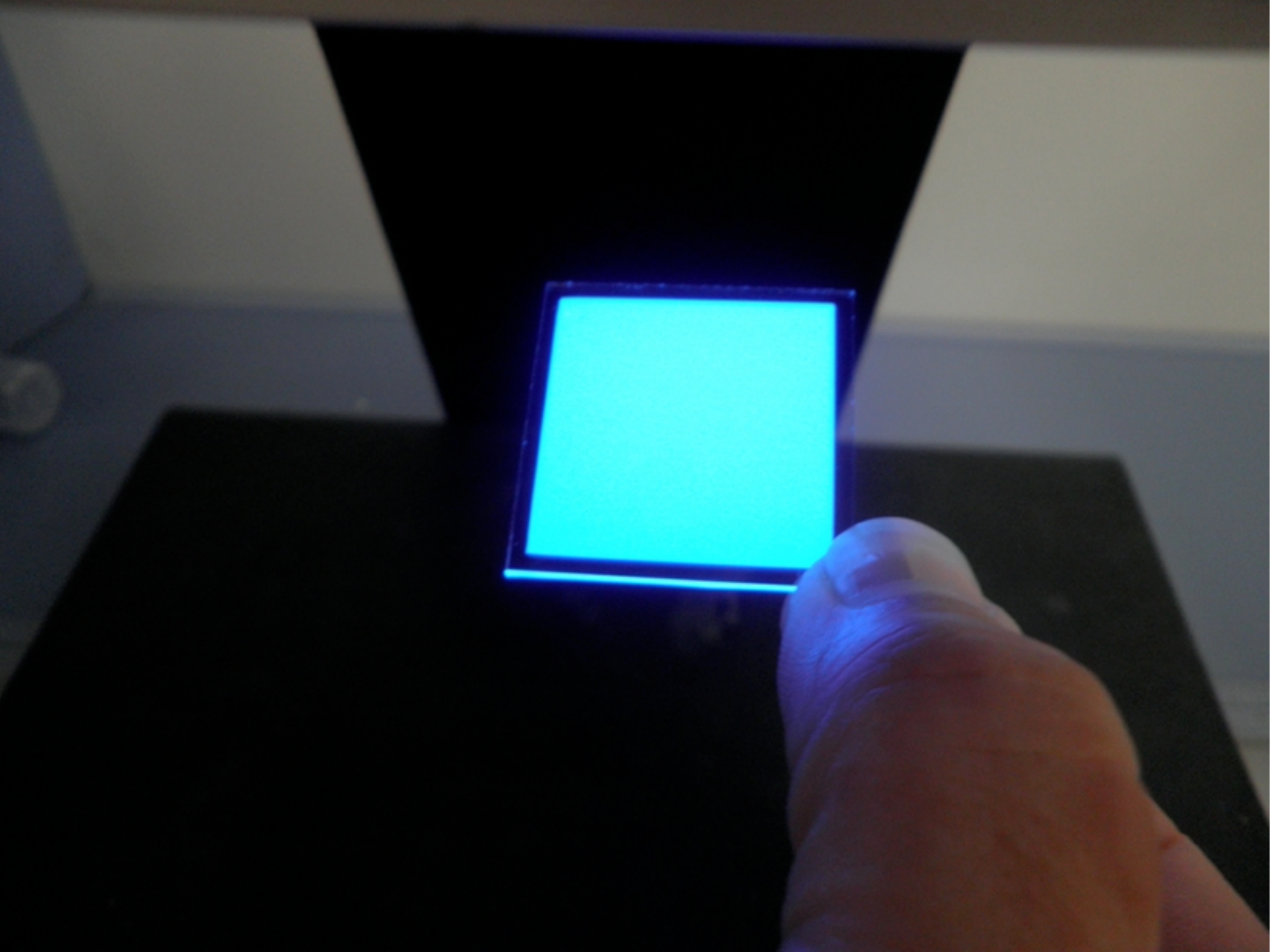}
\includegraphics[width= 6.cm]{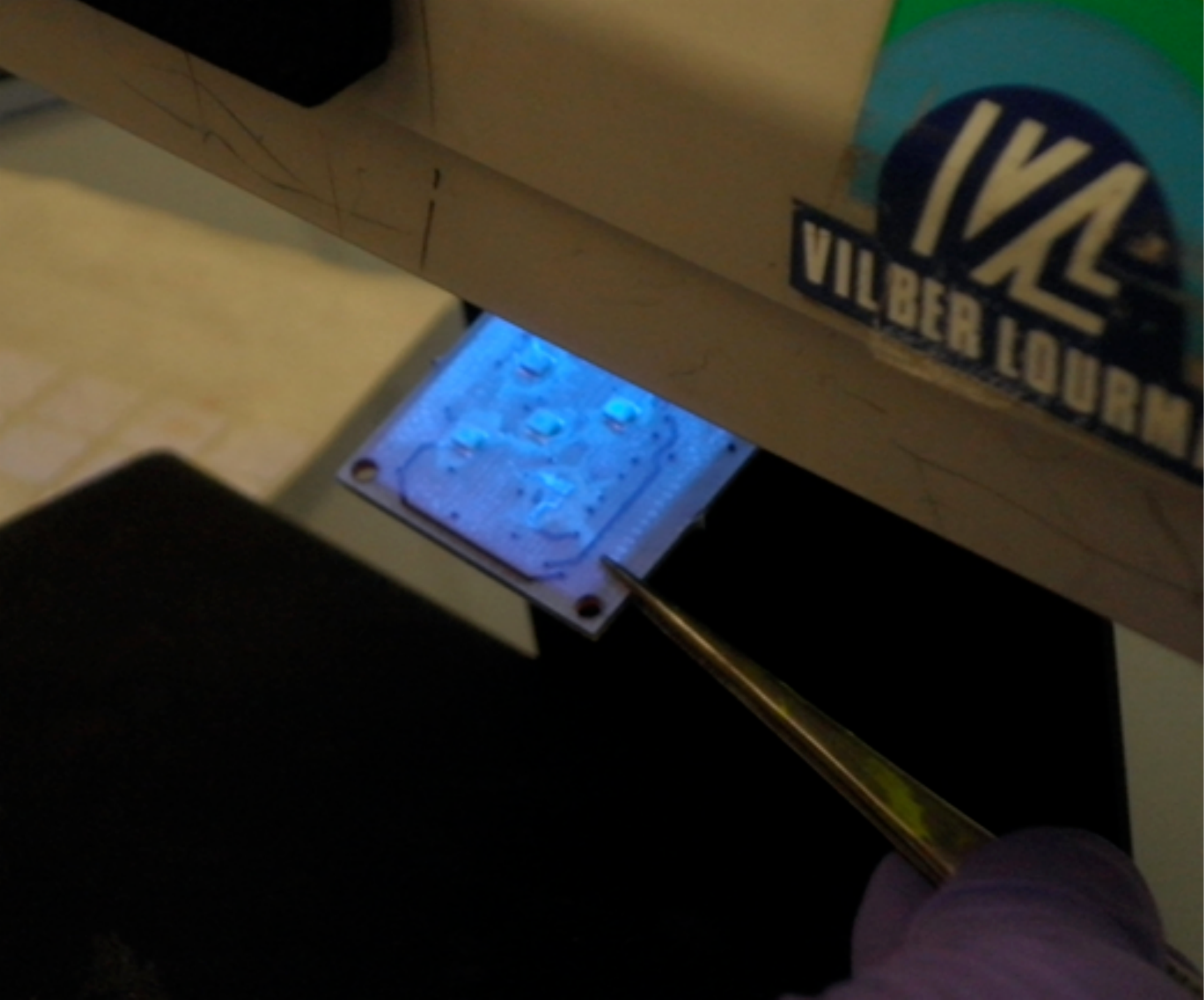}
\caption{Illumination with 240~nm UV light  of a glass-slice (left) and a 5-SiPM board (right) both coated with TPB.} \label{fig:UV_illumination}
\end{figure}

The \emph{photon detection efficiency} (PDE) of the chosen MPPCs peaks in the blue region of the spectrum, and they are not sensitive below 200 nm, where the emission spectrum of xenon lies. Consequently, the DBs will be coated with tetraphenyl butadiene (TPB). The Collaboration has developed a procedure to deposit thin layers of TPB by vacuum evaporation \cite{Alvarez:2012ub}. As an illustration, figure~\ref{fig:UV_illumination} shows a glass-slice (left) and a prototype DB (right) coated with TPB and illuminated with UV light at 240~nm,  clearly re-emitting in the blue.

%%%%%%%%%%
%%%%%%%%%%

\section{The energy plane} \label{sec:EnergyPlane}
The energy measurement in NEXT is provided by the detection of the electroluminescence light by an array of photomultipliers, the \emph{energy plane}, located behind the transparent cathode. Those PMTs will also record the scintillation light that indicates the start of the event. 

\begin{figure}
\centering
\includegraphics[height=6cm]{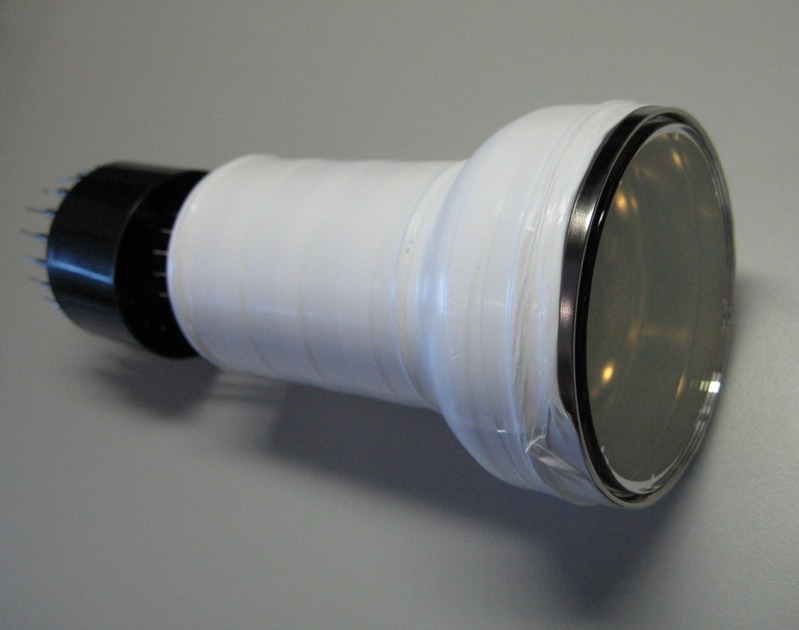}
\caption{The Hamamatsu R11410-10, a 3-inches photomultiplier with high quantum efficiency ($>30\%$) at the xenon scintillation wavelengths and low radioactivity.} \label{fig:PMT}
\end{figure}

A total of 60 Hamamatsu R11410-10 photomultipliers (figure \ref{fig:PMT}) covering 32.5\% of the cathode area constitute the energy plane. This phototube model has been specially developed for radiopure, xenon-based detectors. The manufacturer quoted radioactivity per PMT is 3.3 mBq for the uranium series and 2.3 mBq the thorium series, although independent measurements show even lower activities (see table \ref{tab:RA}). The quantum efficiency of the R11410-10 model is greater than 30\% both in the VUV and around 25\% in the blue region  of the spectrum, and the dark count rate is 2--3 kHz (0.3 photoelectron threshold) at room temperature \cite{Lung:2012pi}.

The PMT coverage is a compromise between the need to collect as much light a possible for energy resolution and the measurement of primary scintillation, and the need to minimize the number of sensors to reduce cost, complexity and radioactivity. Simulations show that this coverage will allow optimal detection of events with energies well below 100 keV in the full chamber range.

%%%%%%%%%%
%%%%%%%%%%

\begin{figure}
\centering
\includegraphics[width=.75\textwidth]{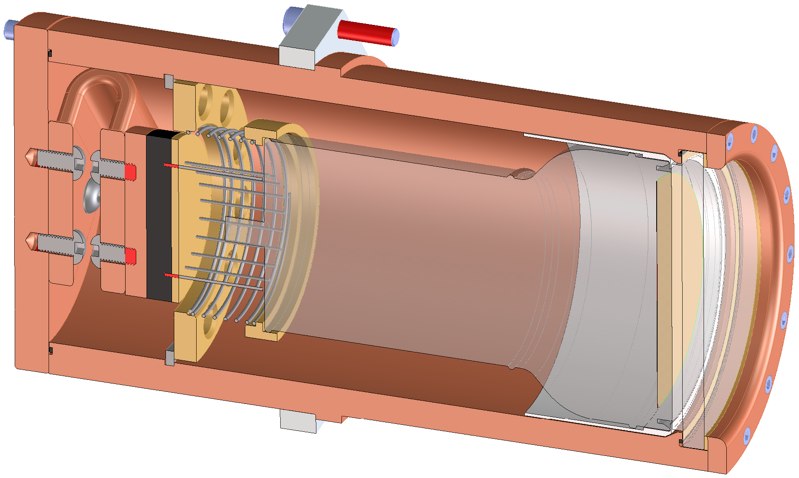}
\caption{A photomultiplier inside its pressure-resistant enclosure.} \label{fig:PmtModule}
\end{figure}

Pressure-resistance tests run by the manufacturer showed that the R11410-10 cannot withstand pressures above 6 atmospheres \cite{Lung:2012pi}. Therefore in NEXT-100 they will be sealed into individual pressure resistant, vacuum tight copper enclosures coupled to sapphire windows (see figure \ref{fig:PmtModule}). The window, 5 mm thick, is secured with a screw-down ring and sealed with an O-ring to the front-end of the enclosure. A similar back-cap of copper seals the back side of the enclosures. The PMT is optically coupled to the window using silicone optical pads of 2--3 mm thickness. A spring on the backside pushes the photomultiplier against the optical pads.

%%%%%%%%%%
%%%%%%%%%%

\begin{figure}
\centering
\includegraphics[width=.8\textwidth]{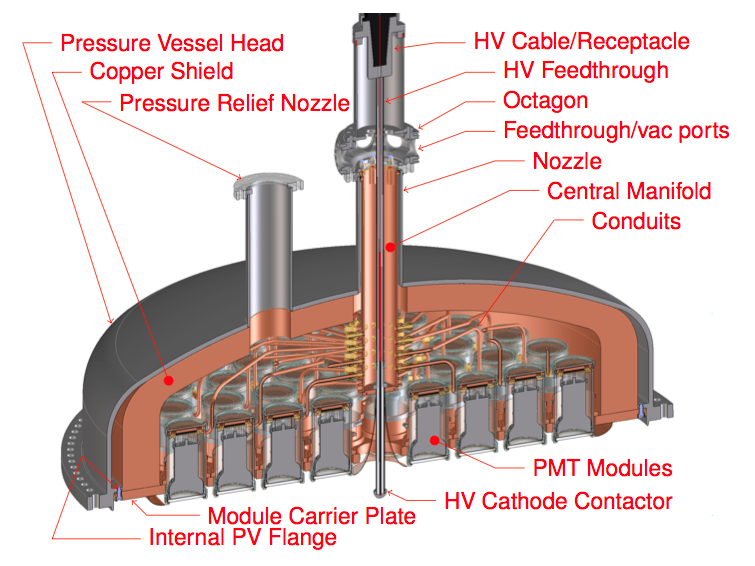}
\caption{The full energy plane of NEXT-100 mounted in the vessel head.} \label{fig:EnergyPlane}
\end{figure}

These PMT modules are all mounted to a common carrier plate that attaches to an internal flange of the pressure vessel head (see figure \ref{fig:EnergyPlane}). The enclosures are all connected via individual pressure-resistant, vacuum-tight tubing conduits to a central manifold, and maintained at vacuum well below the Paschen minimum, avoiding sparks and glow discharge across PMT pins. The PMT cables route through the conduits and the central manifold to a feedthrough in the pressure vessel nozzle. 

%%%%%%%%%%
%%%%%%%%%%

This design requires a vacuum inside the enclosure, so as to detect the presence of any Xe leakage. Without vacuum, the enclosure would eventually pressurize and destroy the PMT. Xenon leakage through seals --- calculated to be no more than 300 ${\rm g}/{\rm year}$ ---will be recovered in a cold trap in the vacuum system (see section \ref{sec:GasSystem}).

\section{Front-end electronics and DAQ} \label{sec:Electronics}
%%%
The NEXT-100 data-acquisition system (DAQ) follows a modular architecture named the Scalable Readout System (SRS), already described in our CDR \cite{Alvarez:2011my}. At the top of the hierarchy, a PC farm running the DAQ software, DATE, receives event data from the DAQ modules via Gigabit Ethernet (GbE) links. The DATE PCs (Local Data Concentrators, LDCs) assemble incoming fragments into sub-events, which are sent to one or more additional PCs (Global Data Concentrators, GDC). The GDCs build complete events and store them to disk for offline analysis.

The DAQ modules used are Front-End Concentrator (FEC) cards, which serve as the generic interface between the DAQ system and application-specific front-end modules. The FEC module can interface different kinds of front-end electronics by using the appropriate plug-in card. The FEC card and the overall SRS concept have been developed within the framework of the CERN RD-51 collaboration. Three different FEC plug-in cards are used in NEXT-100. The cards responsible for the energy plane readout digitization and for the trigger generation are described in section \ref{subsec:fe_energy}, where the energy plane analog front-end is also described. The tracking plane readout, described in section \ref{subsec:fe_tracking}, uses a third type of plug-in card.

%%%%%%%%%%%%%%%%%%%%%%%%%%%%%%%%%%%%%%%%%%%%%%%%%%%%%%%%%%%%
\subsection{Electronics for the energy plane} \label{subsec:fe_energy}
The front-end (FE) electronics for the PMTs in NEXT-100 will be very similar to the system developed for the NEXT-DEMO and NEXT-DBDM prototypes. The first step in the chain is to shape and filter the fast signals produced by the PMTs (less than 5 ns wide) to match the digitizer and eliminate the high frequency noise. An integrator is implemented by simply adding a capacitor and a resistor to the PMT base. The charge integration capacitor shunting the anode stretches the pulse and reduces the primary signal peak voltage accordingly.

Our design uses a single amplification stage based on the fully differential amplifier THS4511, which features low noise (2 ${\rm nV}/\sqrt{\rm Hz}$) and provides enough gain to compensate for the attenuation in the following stage, based on a passive RC filter with a cut frequency of 800 kHz. This filtering produces enough signal stretching to allow acquiring many samples per single photo-electron at 40 MHz. The front-end circuit for NEXT-DEMO was implemented in 7 channel boards and connected via HDMI cables to 12-bit 40-MHz digitizer cards. These digitizers are read out by the FPGA-based DAQ modules (FEC cards) that buffer, format and send event fragments to the DAQ PCs. As the FEC card, also the 16-channel digitizer add-in card has been designed as a joint effort between CERN and the NEXT collaboration within the RD-51 R\&D program. These two cards are edge mounted to form a standard 6U$\times$220 mm eurocard. The energy plane readout system for NEXT-100 will use 4 FEC cards to read 60 PMT channels. 

An additional FEC module with a different plug-in card is used as trigger module. Besides forwarding a common clock and commands to all the DAQ modules, it receives trigger candidates from the DAQ modules, runs a trigger algorithm in the FPGA and distributes a trigger signal. The trigger electronics accepts also external triggers for detector calibration purposes.

%%%%%%%%%%%%%%%%%%%%%%%%%%%%%%%%%%%%%%%%%%%%%%%%%%%%%%%%%%%%
\subsection{Electronics for the tracking plane} \label{subsec:fe_tracking}
The tracking plane will have $\sim$ 7\,000 channels. Passing all those wires across feedthroughs, as it has been done for NEXT-DEMO, is possible but challenging, and probably not optimal. Consequently we are developing a new in-vessel FE electronics system that reduces the total number of feedthroughs to an acceptable level. Here we present the new electronics readout architecture.

\begin{figure}
\centering
\includegraphics[width=0.5\paperwidth]{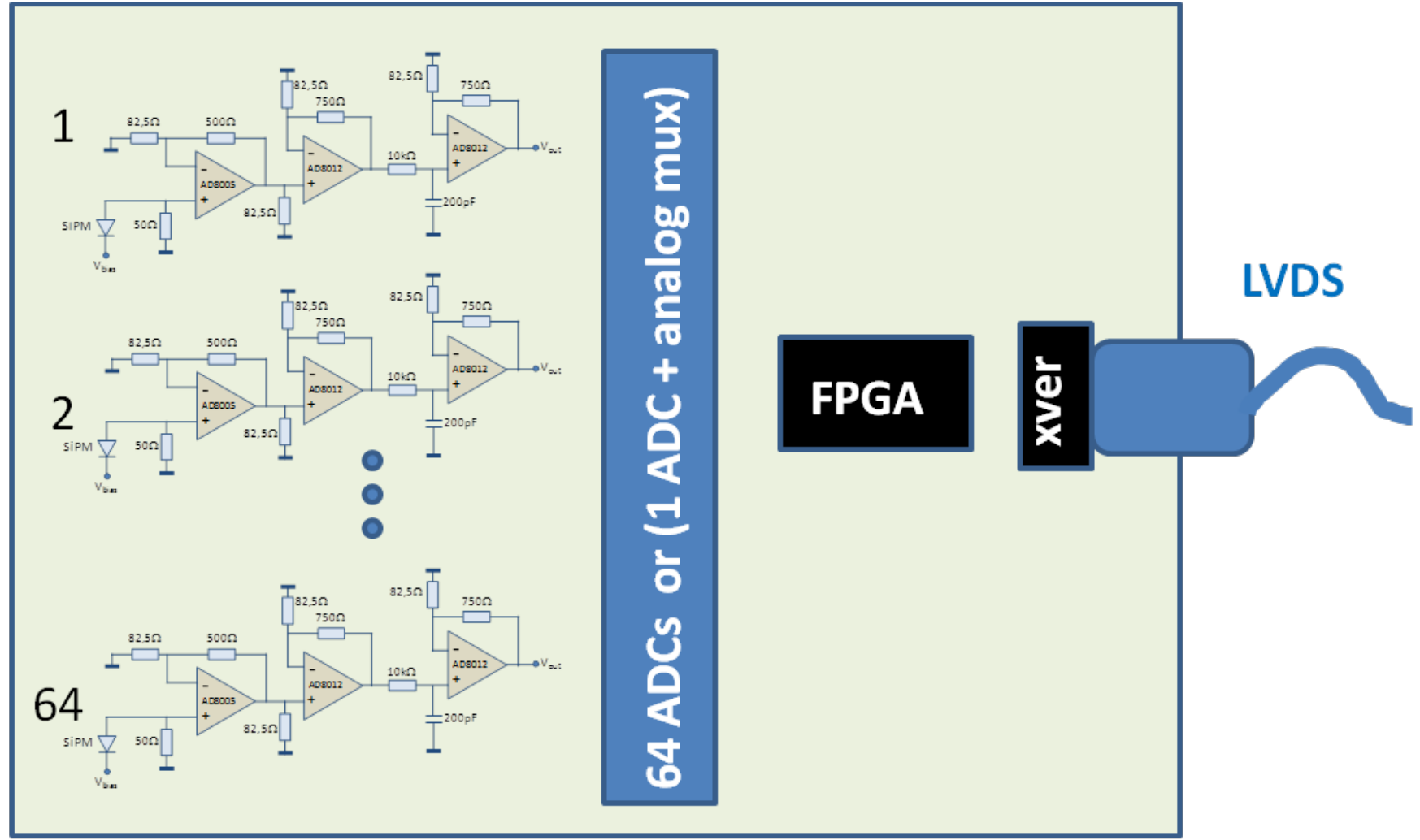}
\caption{\small Functional blocks in the FEB card.}
\label{fig:feb}
\end{figure}

Since the electronics will be inside the PV, it must necessarily be very low power to minimize the heat dissipated inside the vessel. Our design consists of a very simple Front-End Board (FEB, fig.~\ref{fig:feb}) to be placed inside the detector. The 64-ch FEB takes the input of a single DB (transmitted via low-crosstalk kapton ribbon cables) and includes the analog stages, ADC converters, voltage regulators and an FPGA that handles, formats, buffers and transmits data to the outer DAQ. LVDS clock and trigger inputs are also needed. A total of 110 FEBs are required.

%%%%%%%%%%
%%%%%%%%%%

\begin{figure}
\centering
\includegraphics[width=0.5\paperwidth]{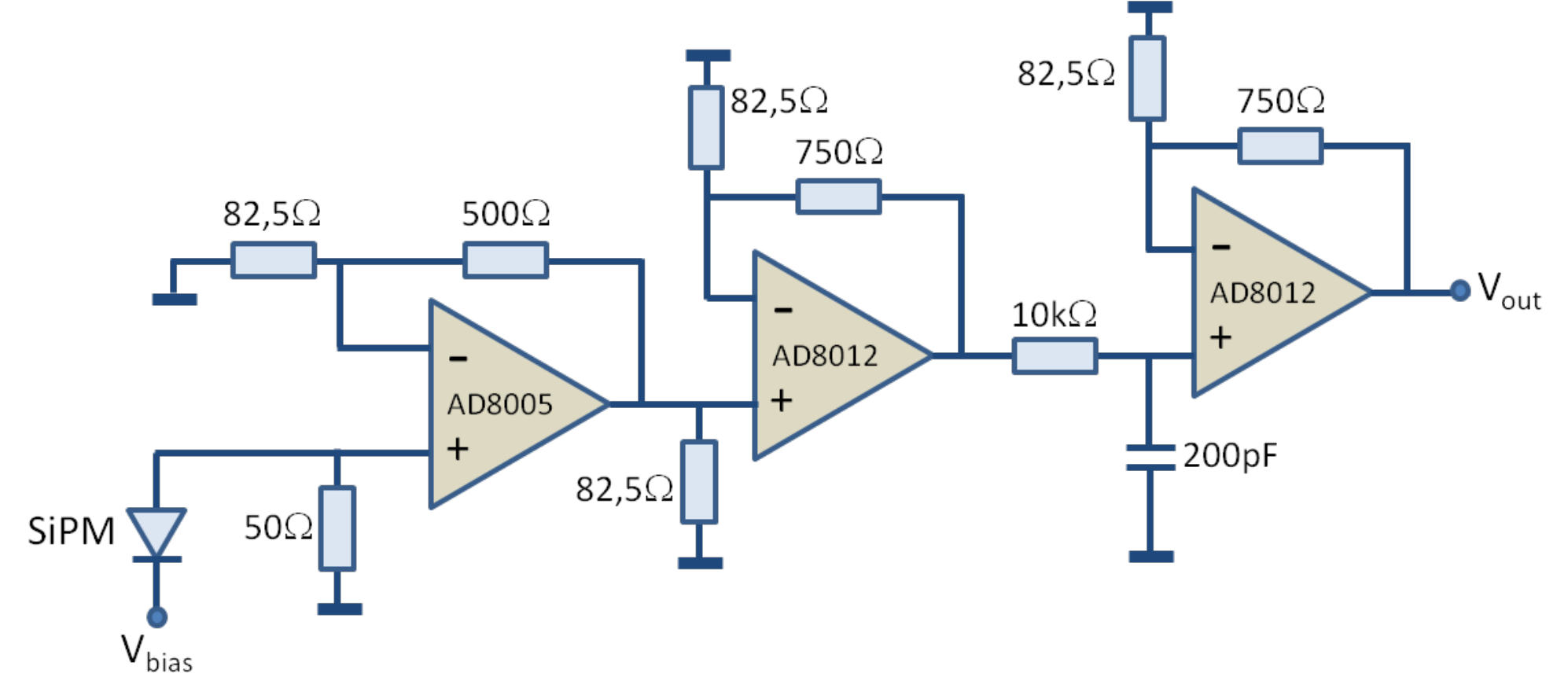}
\caption{This low power amplifier circuit for NEXT-100 features only 30 mW power, 4 mV/pe gain and 1.7 mV rms noise.} \label{fig:analogfe_tracking}
\end{figure}

The low power consumption is achieved by using a passive RC circuit and ultra low power amplifiers, rather than gated integrators and power-hungry devices as was done in NEXT-DEMO. Figure \ref{fig:analogfe_tracking} shows our design for the analog stage. It is a three-stage circuit, with a gain of 10 in each stage. The first two stages are based on the AD8012 (two amplifiers per package, very low noise) and the last one on the AD8005 (ultra low power, 400 $\mu$A quiescent current). The total gain is ($R_t$ is the input termination resistance) $10^3\times R_t=5\times 10^4$, as the first stage is a transimpedance amplifier with gain of $10\times R_t=5\times 10^2$. A passive, 2 $\mu$s time-constant RC circuit (200 pF, 10 k$\Omega$ between the second and the third stage) acts as the circuit integrator. This gain will result in a 1 V output for a 250-pe dynamic range. Total electronic noise in the amplifier circuit is very low according to the simulations: 1.7 mV rms. A preliminary estimation for the power dissipation due to the analog stage yields 30 mW per channel, or 210 W in total. Additional power dissipation in the FEB comes from drop in voltage regulators, FPGA (data handling and multiplexing) and transmission circuits required to reduce the number of feedthroughs in the TPC vessel.  

%%%%%%%%%%
%%%%%%%%%%

For the tracking plane readout digitization two alternatives are currently under study:
\begin{itemize}
\item A single-channel 1 MHz (or 3 MHz) low-power 12-bit ADCs (like AD7476), requiring only two lines (data and chip select) for readout. This solution is used in NEXT-DEMO, and would correspond to a total heat dissipated by the in-vessel electronics of 315 W. 
\item A fast (40 MHz or higher) multi-channel ADC and an analog switch for multiplexing. This solution can lead to a reduced power consumption, though the effect of the noise induced by hundreds of switches inside the vessel has to be studied.
\end{itemize}

FEB size can be 15$\times$15 cm, leaving 3.5 cm$^2$~ board area per channel. This can easily accommodate the three amplifying stages and ADC per channel plus associated SMD passive components in one board side. The FPGA, voltage regulators and I/O connectors can sit in the opposite layer.

In addition to power consumption, another key figure of merit is FEB throughput. Data from the 6\,800 SiPM channels must be sent across the PV. Minimizing the number of vacuum feedthroughs is a must, and this number is directly proportional to the aggregated throughput. To this end, we envisage the FEB readout to be zero suppressed and triggered. 

Zero suppression implies that every microsecond (the ADC sampling period), only channels with a charge readout above an adjustable threshold will send digitized and timestamped data to the DAQ module, where the data are stored in a circular buffer. Raw data mode of operation, where no zero suppression occurs at the online level, will also be supported for testing purposes.

Additionally, FEB data will be read in triggered mode. For a 10 Hz trigger rate and a 2 ms event duration, a triggered FEB readout may further reduce the FEB throughput by a factor of 50 with respect to a continuous readout. In continuous readout mode, no trigger exists at the front-end level, and (zero-suppressed) data are sent continuously to the DAQ module, every microsecond. Once a timestamped trigger arrives to the DAQ module, the right data time interval is read from the DAQ module circular buffer and sent to the DATE online system for event building. In triggered readout mode, on the other hand, FEB data are sent to the DAQ module only in the presence of a trigger. The reduced throughput of the triggered mode comes at the cost of increased complexity, as circular (ring) buffers are needed at the FEB level. Quantitatively, 7\,000 channels produce approximately 20 MByte/event in raw data mode (no zero suppression). A 10 Hz FEB trigger rate implies therefore a 200 MByte/s, or 1.6 Gb/s, throughput. This is an acceptable number.

The necessary FEB circular buffer size has also been estimated, and possible solutions have been identified. One full event (2\,000 samples) requires $2.000\times 64\times 12$ bit, that is 1.5 Mb buffer size. For zero-dead-time operation the buffer must be large enough to be continuously filled. A buffer of about 10 ms is therefore needed, assuming a 200 Mb/s link speed to read the buffer. A conservative estimate for the FEB buffer size is therefore 7.5 Mb. This is feasible with Xilinx Artix FPGAs (XC7A200T), that has an internal 13.5 Mb memory. A trade-off between readout speed and buffer size must be found in order to minimize cost and power: data links must a have a low speed for the sake of reduced power consumption, cheaper electronics, reduced FEB noise and enhanced signal integrity. Careful evaluation of copper and optical link solutions will provide the right compromise between buffer size, power dissipation, reliability and cost.

The number of FECs and of Local Data Concentrator (LDC) PCs for the tracking partition of the DAQ is determined by the tracking plane throughput and by the speed of the links (from the in-vessel electronics to the FEC card, and from the FEC card to the LDC PC). As discussed above, the tracking plane will produce 1.6 Gb/s data at most (10 Hz triggered mode, no zero suppression). Assuming 400 Mb/s as a comfortable working point for the gigabit Ethernet links between the FECs and the LDCs, 4 LDCs are required. Assuming 200 Mb/s link speed (LVDS over copper) from the in-vessel electronics to the DAQ (the same speed and technology used in NEXT-DEMO for the SiPM plane readout), the existing 16-link LVDS add-in card for the FEC module can be used. The 110 links coming from the vessel require then 110/16=7 FEC cards. We therefore need 7 FECs and 4 LDCs in the tracking DAQ partition. This is approximately the size of the full NEXT-DEMO DAQ system.

%%%%%%%%%%%%%%%%%%%%%%%%%%%%%%%%%%%%%%%%%%%%%%%%%%%%%%%%%%%%
\subsection{Trigger electronics} \label{subsec:fe_trigger}
As it was previously mentioned, one additional FEC module with a different plug-in card is used as trigger module. This plug-in card (named CDTC16 from Clock, Data, Trigger and Control) has 16 RJ-45 connectors and 4 LVDS pairs on each connector. Every readout FEC card (for both energy and tracking planes) is connected to the trigger module via a 4-pair CAT6 network cable to the trigger module.

One LVDS pair is used to distribute a common clock to all the readout FEC. A second LVDS pair is used to send common trigger, synchronization and configuration commands to all readout FECs. Two pairs carry trigger candidates from each readout FEC to the trigger module, where a trigger decision is taken. Accepted triggers are timestamped and sent to all readout FECs. Additionally, the trigger module can send the DAQ information related to the trigger type and conditions that produces the trigger for offline analysis purposes.

The trigger algorithm runs on the FEC FPGA. This allows very low latencies (the algorithm runs on hardware, not on software) and virtually any kind of trigger one can think of. As an example, three basic trigger types have been initially defined: (1) External trigger (NIM or LVDS pulse); (2) Internal trigger (at least N --- programmable number --- PMT channels above a programmable threshold are in coincidence in a programmable time window, using only a programmable set of PMTs -to ignore noisy or damaged devices); and (3) Mixed trigger (internal trigger gated by an external trigger). More complex trigger types can be defined with the only limitation of the FPGA logic resources.

The described trigger architecture has been implemented and tested in NEXT-DEMO and will be also used in NEXT-100. New or modified trigger algorithms will be developed for NEXT-100.

\section{Shielding} \label{sec:Shielding}
\begin{figure}
\centering
\includegraphics[width=0.6\textwidth]{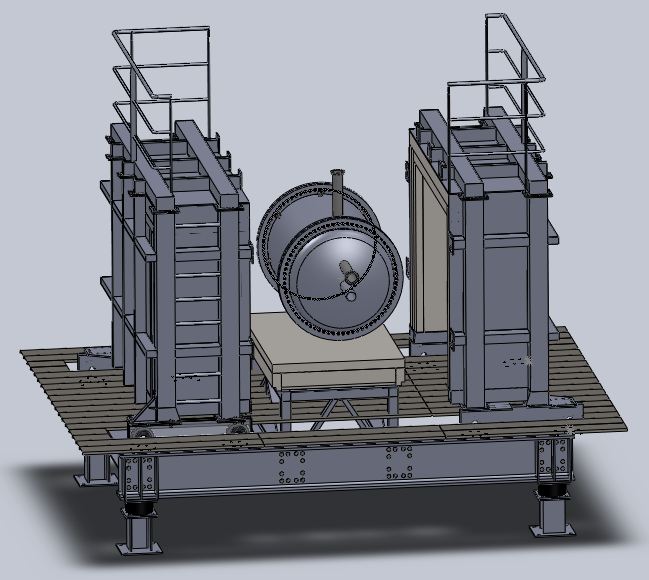}
\caption{Drawing of the NEXT-100 lead castle shield in its open configuration.}\label{fig:Shielding}
\end{figure}

To shield NEXT-100 from the external flux of high-energy gamma rays a relatively simple lead castle, shown in figure \ref{fig:Shielding}, has been chosen, mostly due to its simplicity and cost-effectiveness. The lead wall has a thickness of 20 cm and is made of layers of staggered lead bricks held with a steel structure. The lead bricks have standard dimensions ($200\times100\times50$ mm$^3$), and, by requirement, an activity in uranium and thorium lower than $0.4\ {\rm mBq}/{\rm kg}$. 

%%%%%%%%%%
%%%%%%%%%%

The lead castle is made of two halves mounted on a system of wheels that move on rails with the help of an electric engine. The movable castle has an open and a closed position. The former is used for the installation and service of the pressure vessel; the latter position is used in normal operation. A lock system fixes the castle to the floor in any of the two configurations to avoid accidental displacements.

\section{Gas system} \label{sec:GasSystem}
\begin{figure}
\centering
\includegraphics[width=\textwidth]{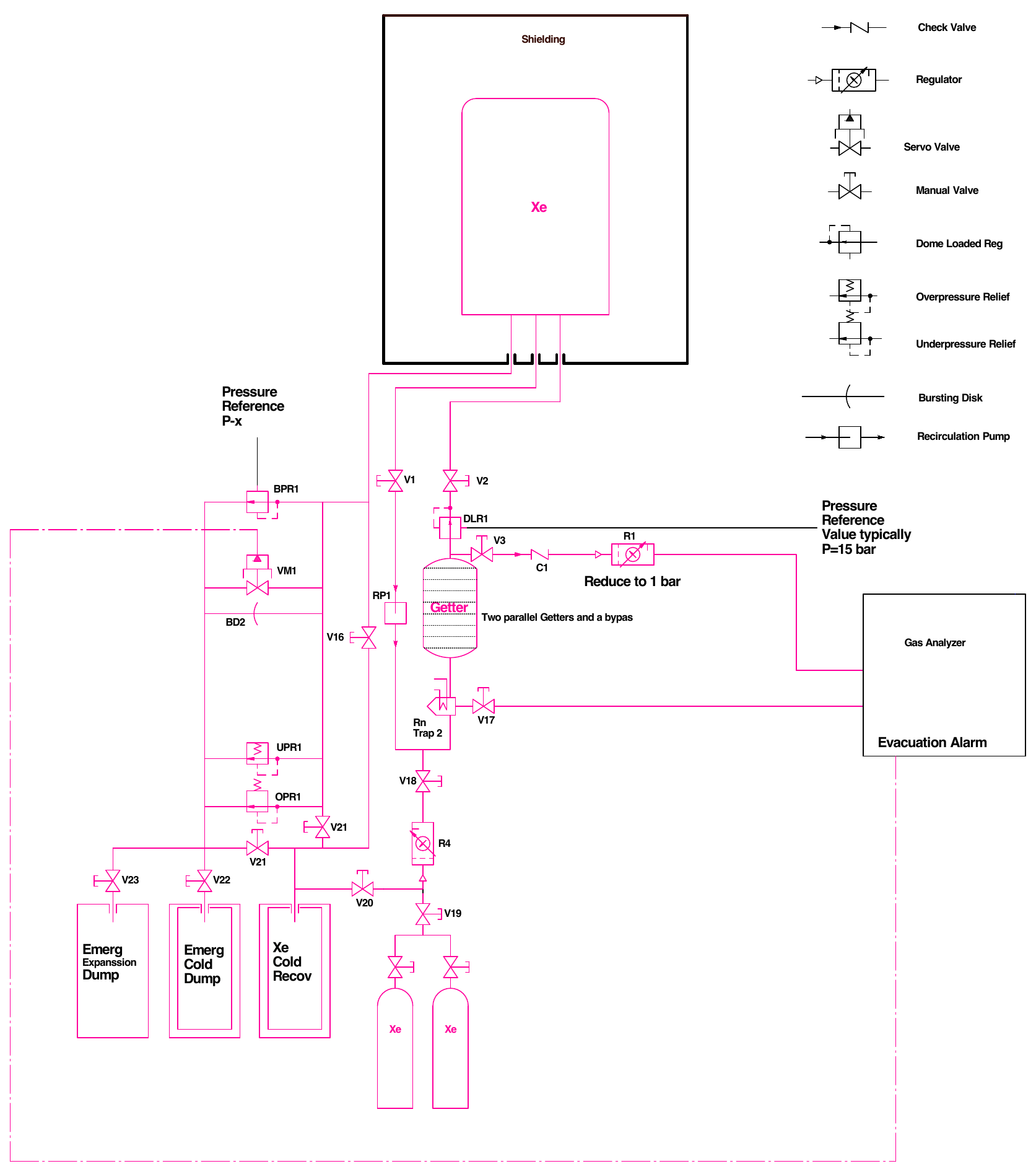} 
\caption{Schematic of the NEXT-100 gas system.} \label{fig:GasSystemSchem}
\end{figure}

The gas system must be capable of pressurizing, circulating, purifying, and depressurizing the NEXT-100 detector with xenon, argon and possibly other gases with negligible loss and without damage to the detector. In particular, the probability of any substantial loss of the very expensive enriched xenon (EXe) must be minimized. The general schematic of the gas system is given in figure \ref{fig:GasSystemSchem} (the re-circulation compressor, vacuum pump and cold traps are not shown). A list of requirements, in approximate decreasing order of importance, considered during the design is given below:
\begin{enumerate}
\item Pressurize vessel, from vacuum to 15 bar (absolute).
\item Depressurize vessel to closed reclamation system, 15 bar to 1 bar (absolute), on fault, in 10 seconds maximum.
\item Depressurize vessel to closed reclamation system, 15 bar to 1 bar (absolute), in normal operation, in 1 hour maximum.
\item Pressure relief (vent to closed reclamation system) for fire or other emergency condition.
\item Maximum leakage of EXe through seals (total combined): $100\ {\rm g}/{\rm year}$.
\item Maximum loss of EXe to atmosphere: $10\ {\rm g}/{\rm year}$.
\item Accomodate a range of gasses, including Ar and N$_2$\label{gases}. 
\item Circulate all gasses through the detector at a maximum rate of 200 standard liters per minute (slpm) in axial flow pattern.
\item Purify EXe continuously. Purity requirements: $<1$ ppb O$_2$, CO$_2$, N$_2$, CH$_4$.
\end{enumerate}

%%%%%%%%%%
%%%%%%%%%%

To insure the cleanliness of the chamber and the gas system prior to the introduction of the xenon gas, both need to be vacuum evacuated to as low pressure as possible. A reasonably good vacuum is in the range of $10^{-4}$ to $10^{-5}$ mbar. To achieve this, 
a turbo-molecular pump station will be directly connected as close as possible to the NEXT-100 vessel through a large conductance valve rated for vacuum and pressure. However, many internal structures of the NEXT-100 detector will not allow good conductance for vacuum evacuation. Therefore, instead of evacuating the system from a single point, the vacuum manifold will be connected to several points simultaneously. Also, flushing with argon several times might help in the cleaning process. Finally, continuous gas re-circulation through the getters will clean the gas system.

The most vulnerable component of the gas system is the re-circulation compressor, that must have sufficient redundancy to minimize the probability of failure and leakage. The Collaboration has chosen a compressor manufactured by \textsc{sera ComPress GmbH}. This compressor is made with metal-to-metal seals on all the wetted surfaces. The gas is moved through the system by a triple stainless steel diaphragm. Between each of the diaphragms there is a sniffer port to monitor for gas leakages. In the event of a leakage, automatic emergency shutdown can be initiated.

\textsc{MicroTorr} model MC4500-902FV from SAES has been chosen as the purification filter for the xenon gas. Capable of removing electronegative impurities to less than 1 ppb, the chosen model has a nominal flow rate of 200 standard liters per minute, well in excess of the required flow rates for NEXT-100, thus offering sufficient spare capacity. The gas system will contain two such getters in parallel with a bypass. The second spare getter is placed in parallel allowing uninterrupted running in the event of accidental contamination of one of the getters. Also, the ability to bypass the getters will allow the testing of the purification of the gas and aid in diagnostic and monitoring of the gas system. While cold getter technology is capable of reaching the required purity levels in water and oxygen, a hot getter can also remove nitrogen and methane. In that regard, we foresee to upgrade to hot getters, such as the model PS4-MT15 from SAES, for the enriched xenon run.

An automatic recovery system of the expensive EXe will be needed to evacuate the chamber in case of an emergency condition. A 30-m$^3$ expansion tank will be placed inside the laboratory to quickly reduce the gas pressure in the system. Additionally, we will implement a similar solution to that proposed by the LUX collaboration, where a chamber permanently cooled by liquid nitrogen will be used. Two primary conditions to trigger automatic evacuation are foreseen:
\begin{itemize}
\item An over-pressure, that can potentially cause an explosion. Because the gas system for NEXT-100 will be operated in a closed mode the overpressure condition could occur only under two possible scenarios: a problem during  the filling stage of the operation or a thermal expansion of the gas due to laboratory fire. In the case of overpressure an electromechanical valve, activated by a pressure switch,  will open a pipe from the chamber to a permanently cold recovery vessel. This will then cryo-pump xenon into the recovery vessel, causing the gas to freeze in the recovery tank. In the event of the electromechanical valve failing, a mechanical spring-loaded relief valve, mounted in parallel to the electromechanical valve, would open and allow the xenon to be collected in the recovery vessel. A bursting disk will also be mounted in parallel to the electromechanical  and spring-loaded valves  as a final safety feature.
\item An under--pressure, indicating a leak in the system. Such condition  would require evacuation of the chamber to prevent losses of gas. If this happens an electromechanical valve sensing under-pressure will open and evacuate the xenon into the recovery vessel.
\end{itemize}

We have also considered the scenario in which xenon could leak through some of the photomultipliers enclosures (leaking can). If this happens the use of a cold trap would permit to recover the gas. 

\section{NEXT-100 at the LSC}\label{sec:Infrastructure}
\begin{figure}
\centering
\includegraphics[height=7cm]{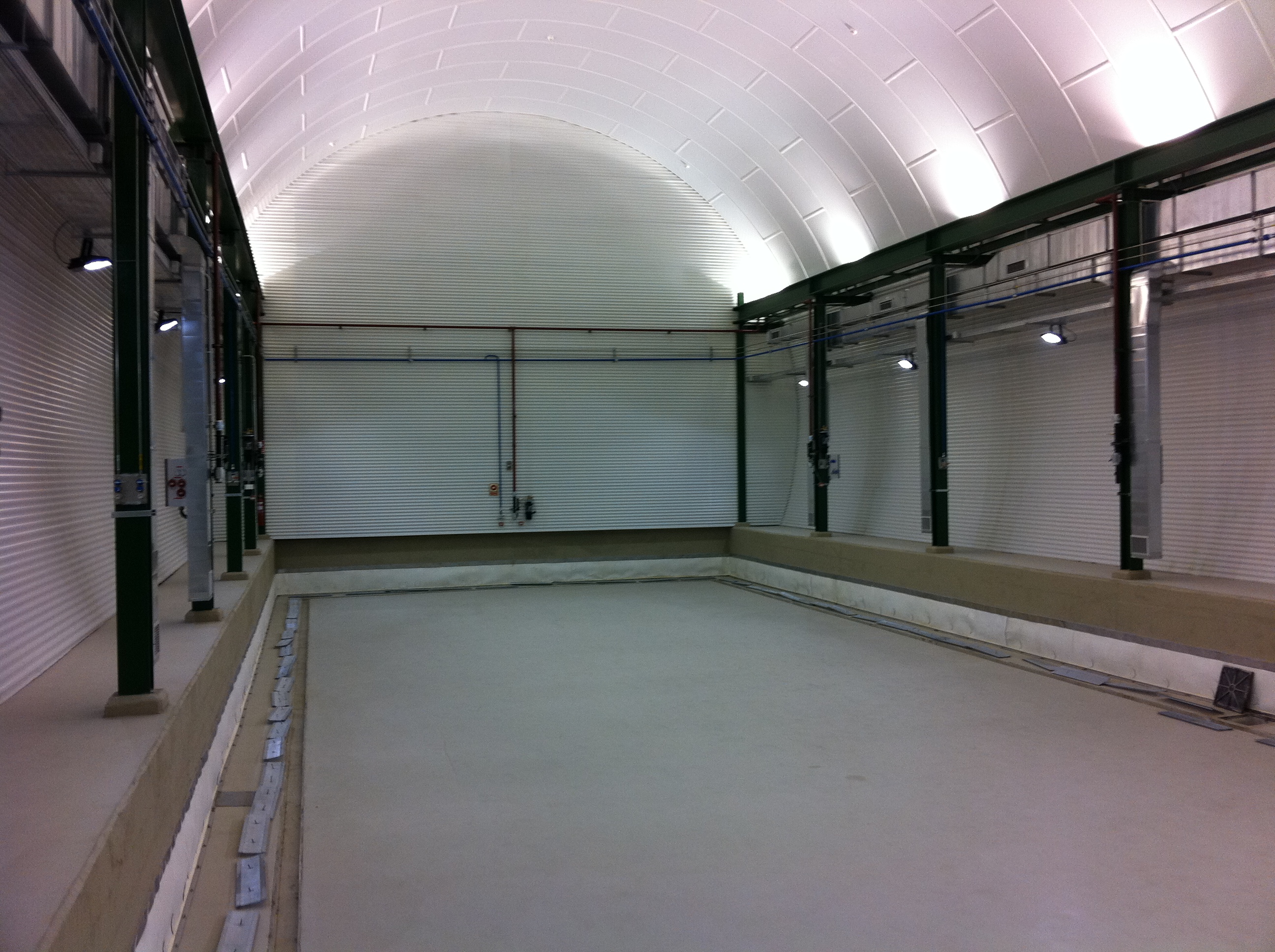}
\caption{View of Hall A of the Laboratorio Subterr\'aneo de Canfranc prior to any equipment installation.} \label{fig:LSC_Hall_A}
\end{figure}

The Collaboration intends to commence the installation at the LSC of NEXT-100 and its ancillary systems in the second quarter of 2012. Figure \ref{fig:LSC_Hall_A} shows an image of Hall A, future location of NEXT-100. The pool-like structure is intended to be a catchment reservoir to hold xenon or argon --- a liquid-argon experiment, ArDM, will be neighboring NEXT-100 in Hall A --- gas in the event of a catastrophic leak. Therefore, for reasons of safety all experiments must preclude any personnel working below the level of the top of the catchment reservoir.

%%%%%%%%%%
%%%%%%%%%%

An elevated working platform will be built prior to the installation of NEXT-100. It is designed to stand a uniform load of $1500\ {\rm kg}/{\rm m}^2$ and a concentrated load of $200\ {\rm kg}/{\rm m}^2$. It is anchored to the hall ground and walls. The platform floor tiles are made of galvanized steel and have standard dimension to minimize cost. 

Due to the mild seismic activity of the part of the Pyrenees where the LSC is located, a comprehensive seismic study has been conducted as part of the project risk analysis. As a result, an anti-seismic structure that will hold both pressure vessel and shielding has been designed. This structure will be anchored directly to the ground and independent of the working platform (see section \ref{sec:Infrastructure}) to allow seismic displacements in the event of an earthquake. 

\begin{figure}
\centering
\includegraphics[height=5cm]{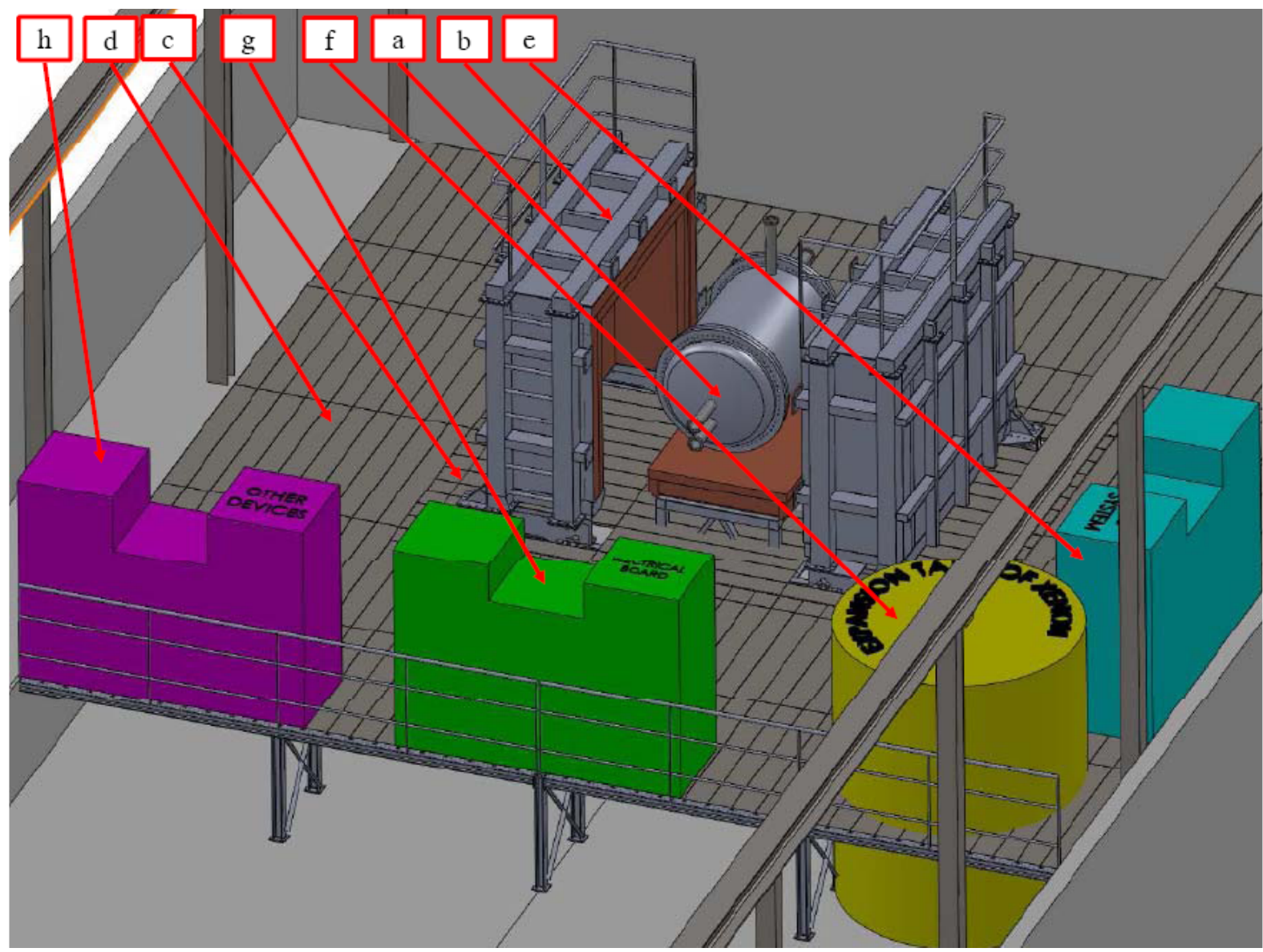}
\includegraphics[height=5cm]{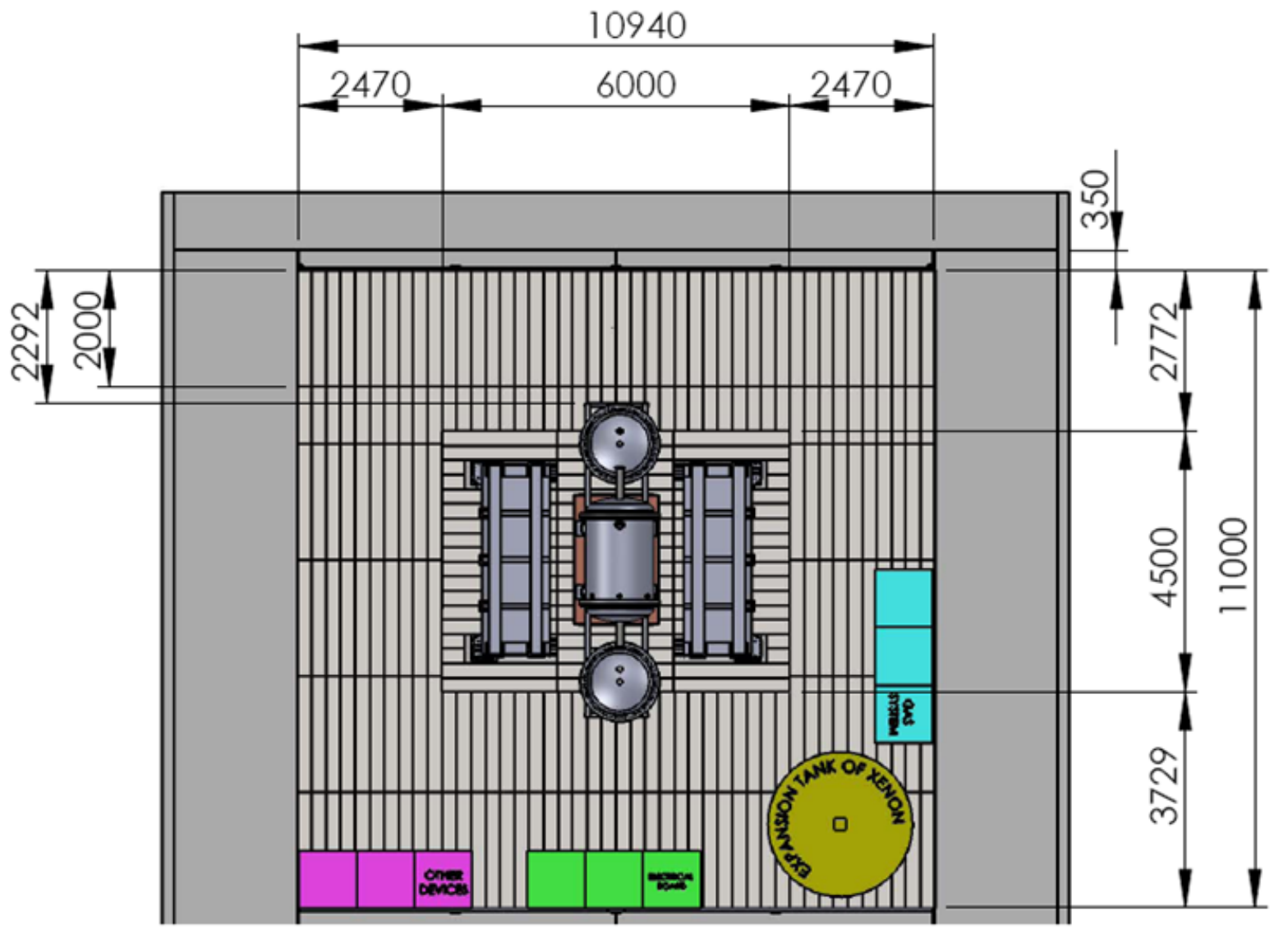}
\caption{Left: Intended location of the components  and subsystems for the operation of NEXT-100 on the working platform: (a) NEXT-100; (b) the lead castle shield in its open configuration; (c) seismic platform; (d) working platform; (e) gas purification system; (f) emergency gas vent tank; (g) data acquisition system; (h) other systems. Right: Top view showing the dimensions of the working platform.} \label{fig:Infrastructure}
\end{figure}

Figure \ref{fig:Infrastructure} shows the placement of NEXT-100 and components on the platform as well as the dimensions. Work is underway, in coordination with LSC staff to refine and complete the design of each relevant element, and the integration of all the systems. 

%%%%%%%%%%
%%%%%%%%%%

\section{Background model}\label{sec:Background}
%%%%%%%%%%%%%%%%%%%%%%%%%%%%%%%%%%%%%%%%%%%%%%%%%%%%%%%%%%%%
\subsection{Sources of background in NEXT} \label{subsec:sourcesbg}
NEXT has two powerful handles to distinguish signal from background:
\begin{itemize}
\item \emph{Energy resolution}: Signal events have all the same energy. Selecting only the events in the energy region around \Qbb\ defined by the resolution eliminates most of the spurious activity in the detector.
\item \emph{Event topology}: Signal events appear uniformly distributed in the source (i.e., the enriched xenon) and have a distinctive topology (a twisted track with blobs in both ends, see figure \ref{fig:track}). Requiring signal events to be strictly contained in the active volume of the chamber eliminates essentially all charged backgrounds entering the detector. Confined tracks generated by neutral particles, like high-energy, gammas can be suppressed by pattern recognition.
\end{itemize}
The relevance of a background source depends therefore on its probability of generating a signal-like track in the active volume with energy around \Qbb.

The \bbonu\ peak of \XE\ is located in the energy region of the naturally-occurring radioactive processes. The half-life of the parents of the natural decay chains, of the order of the age of the universe, is, however, very short compared to the desired half-life sensitivity of the new \bbonu\ experiments ($>10^{25}$ years). For that reason, even small traces of these nuclides create notable background rates, and screening and selection of radiopure construction materials is mandatory (see section \ref{subsec:radiopurity}). The $Q$-value of \XE\ (2458 keV), in particular, is located between the photoelectric peaks of the high-energy, de-excitation gammas emitted following the $\beta$-decays of \BI, from the uranium series, and \TL, from the thorium series.

The daughter of \TL, \Pb, emits a de-excitation photon of 2614 keV with a 100\% intensity. The Compton edge of this gamma is at 2382 keV, well below \Qbb. However, the scattered gamma can interact and produce other electron tracks close enough to the initial Compton electron so they are reconstructed as a single object falling in the energy region of interest (ROI). Photoelectric electrons are produced above the ROI but can loose energy via bremsstrahlung and populate the window, in case the emitted photons escape out of the detector. Pair-creation events are not able to produce single-track events in the ROI. 

After the decay of \BI, $^{214}$Po emits a number of de-excitation gammas with energies above 2.3 MeV. The gamma line at 2448 keV, of intensity 1.57\%, is very close to \Qbb. The gamma lines above \Qbb\ have low intensity (below 0.1\%), but their Compton spectra can produce background tracks in the ROI.

Radon constitutes a dangerous source of background due to the radioactive isotopes $^{222}$Rn (half-life of 3.8\,d) from the $^{238}$U chain and $^{220}$Rn (half-life of 55\,s) from the $^{232}$Th chain. As a gas, it diffuses into the air and can enter the detector. \BI\ is a decay product of $^{222}$Rn, and \TL\ a decay product of $^{220}$Rn. In both cases, radon undergoes an alpha decay into polonium, producing a negative ion which is drifted towards the anode by the electric field of the TPC.  As a consequence, $^{214}$Bi and $^{208}$Tl contaminations can be assumed to be deposited on the anode surface. The tracking capabilities of NEXT allow the rejection of most of these events. Additionally, radon may be eliminated from the TPC gas mixture by recirculation through appropriate filters. Also, radon can be suppressed in the vicinity of the detector by continuously flushing nitrogen, for instance.

The rock of the laboratory itself is a rather intense source of high-energy gammas from the radioactive decay chains. This flux of photons has been measured \cite{CanfrancFlux}:
\begin{itemize}
\item $0.71 \pm 0.12~{\gamma/\mathrm{cm}^2/\mathrm{s}}$~from the  $^{238}$U chain;
\item $0.85 \pm 0.07~{\gamma/\mathrm{cm}^2/\mathrm{s}}$~from the $^{232}$Th chain.
\end{itemize}
These fluxes include all the gamma lines in each chain. The flux corresponding to the \TL\ line at 2614 keV and the flux corresponding to the \BI\ line at 1764 keV were also measured separately (from the latter it is possible to deduce the flux corresponding to the 2448 keV line). The results are
\begin{itemize}
\item $0.13 \pm 0.01~{\gamma/\mathrm{cm}^2/\mathrm{s}}$~from the \TL\ line;
\item $0.006 \pm 0.001~{\gamma/\mathrm{cm}^2/\mathrm{s}}$~from the \BI\ line at 2448 keV. 
\end{itemize}

Cosmic particles can also affect our experiment by producing high-energy photons or activating materials. Muons are the only surviving cosmic ray particles deep underground, but their interactions with the rock produce neutrons. However, neutron-induced backgrounds do not appear to be significant for our experiment given the tracking capabilities of the detector.

%%%%%%%%%%%%%%%%%%%%%%%%%%%%%%%%%%%%%%%%%%%%%%%%%%%%%%%%%%%%
\subsection{Radioactive budget of NEXT-100} \label{subsec:radiopurity}
Information on radiopurity of the materials expected to be used in
the construction of NEXT-100 has been compiled, performing specific
measurements and also examining data from the literature for materials
not yet screened. In this executive summary we present a brief summary, shown in table \ref{tab:RA}, of our detailed database. 

%%%%%%%%%%
\begin{table}
\caption{Activity (in ${\rm mBq}/{\rm kg}$) of the most relevant materials used in NEXT. Activities for the lower part of the uranium chain, starting at $^{226}$Ra, have been quoted when possible.} \label{tab:RA}
\begin{center}
\begin{tabular}{llcll}
\toprule
Material & Subsystem & Method/Ref. &$^{238}$U & $^{232}$Th  \\ \midrule
Lead, from Cometa & Shielding & GDMS & 0.37 & 0.07 \\
Copper, from Luvata & ICS & GDMS & $<0.012$ & $<0.004$ \\
Steel (316Ti) & PV & \cite{Aprile:2011ru} & $<1.9$ & $<1$ \\
Bolts Inconel 718 & PV & Ge LSC & $<5.6$ & $<4.6$ \\
Bolts Inconel 625 & PV & Ge LSC & $<1.8$ & $<2.0$ \\
PEEK, from Sanmetal & FC/EP/TP & Ge Unizar & 36.3 & 11.7 \\
Capacitors (Tantalum) & FC/EP/TP & \cite{ILI1} & 320 & 1230 \\
SMD Resistors, Finechem (per pc) & FC & Ge Unizar & 0.022 & <0.048 \\
Polyethylene & FC & \cite{Aprile:2011ru} & 0.23 & <0.14 \\
TTX & FC & \cite{BOC09}& 12.4 & <1.6 \\
TPB & FC/EP/TP & \cite{SNOrp} & 1.63 & 0.47 \\
PTFE (Teflon) & EP/TP/DB & \cite{BUD09} & 0.025 & 0.031 \\
PMT (R11410-MOD per pc) & EP & \cite{Aprile:2011ru} & $< 2.5$ & $< 2.5$ \\
PMT (R11410-MOD per pc) & EP & \cite{FAH11}& $< 0.4$ & $< 0.3$ \\
Sapphire window & EP & \cite{LEO08}& <0.31 & 0.12 \\
CUFLON & TP & \cite{NIS09}& 0.36 & 0.28 \\
Kapton cable & TP/EP & \cite{Aprile:2011ru} & <11 & <11 \\
\bottomrule
\end{tabular}
\end{center}
\end{table} 
%%%%%%%%%%

%%%%%%%%%%%%%%%%%%%%%%%%%%%%%%%%%%%%%%%%%%%%%%%%%%%%%%%%%%%%
\subsection{Expected background rate in NEXT-100} \label{subsec:bgrate}
As explained above, electron tracks generated in the active volume by the high-energy gammas emitted in the decays of \TL\ and \BI\ are the main background in NEXT. Therefore, in order to estimate the background rate to be expected in NEXT-100, large samples of \TL\ and \BI\ events were generated in our detector simulation as emanating from the different subsystems, and normalized according to the activities shown in table \ref{tab:RA}. 

Notice that since all background sources are external to the active volume of the detector (the contamination of the enriched xenon is negligible), the background rate in NEXT-100 can be computed as the product of two quantities: the flux of high-energy gammas through the surface of the active volume, and the probability of a background event to be selected as a signal candidate (in the following, the \emph{rejection factor}). The first quantity lets us compare the relative importance of the different subsystems in the radioactive budget of NEXT-100, while the second allows us to better understand the background rejection capabilities of the detector. 

Our first step, therefore, is to quantify via simulation the number of high-energy gammas reaching the active volume. NEXT-100 has the structure of a Matryoshka (a Russian nesting doll). The flux of high-energy gammas emanating from the LSC walls is drastically attenuated by the lead castle (LC), and the residual flux together with that emitted by the lead itself and the materials of the pressure vessel (PV) are further attenuated by the inner copper shielding (ICS). The ICS also attenuates the flux emitted by the tracking plane front-end electronics, which sit behind it. To this, we need to add the contribution from the innermost elements --- sensor planes and field cage --- that are not shielded by the ICS. Table \ref{tab.gtp} summarizes the contributions to the gamma flux from Matryoshka and inner elements. Overall, $3.7\times10^6$ gammas/year from \BI\ and $1.3\times10^6$ gammas/year from \TL\ will reach the active volume. The contributions to this radioactive budget from the different subsystems of the detector are very similar. An important consequence of this is that a substantial reduction of the radioactive budget can only be achieved if it happens in all the subsystems at the same time. In its current configuration the NEXT-100 radioactive budget is quite balanced. 

%%%%%%%%%%%
%\begin{table}
%\caption{Number of high-energy gammas per year outgoing each shell of the Matryoshka. The numbers in bold font are the residual photons after all the shells.}
%\label{tab.matrioska}
%\begin{center}
%\begin{tabular}{lcccccc}
%\toprule
%& \multicolumn{2}{c}{from previous vol after} & \multicolumn{2}{c}{from current vol} &  \multicolumn{2}{c}{total} \\ 
%& \multicolumn{2}{c}{passing through current vol} & \multicolumn{2}{c}{(self shielded)} &  \multicolumn{2}{c}{} \\ \midrule
%  & \BI\ & \TL\  & \BI\ & \TL\ & \BI\ & \TL\ \\ \midrule
%LC & $8.8\times 10^{4}$ & $1.9\times 10^{6}$& $3.2\times 10^{7}$ & $2.1\times 10^{6}$  & $3.2 \times10^{7}$ & $4.0\times 10^{6}$  \\
%PV & $1.2 \times10^{7}$ & $1.6\times 10^{6}$ & $4.5\times 10^{7}$ & $7.8\times 10^{6}$ & $6.4\times 10^{7}$ & $10.5\times 10^{6}$ \\
%ICS & $5.6\times 10^{5}$ & $9.8\times 10^{4}$ & $5.2\times 10^{5}$ & $5.9\times 10^{4}$ & $\mathbf{1.0\times 10^{6}}$ & $\mathbf{1.6\times 10^{5}}$ \\
%\bottomrule
%\end{tabular}  
%\end{center}
%\end{table}
%%%%%%%%%%%

%%%%%%%%%%
\begin{table}
\caption{Number of high-energy gammas per year emitted by the different subsystems of NEXT-100 and reaching the active volume of the detector.}
\label{tab.gtp}
\begin{center}
\begin{tabular}{lrr}
\toprule
 & \BI\ & \TL\ \\ 
 & \multicolumn{2}{c}{[$10^{5}\ \gamma/{\rm year}$]} \\ \midrule
Matryoshka & 10.0 & 1.6 \\
Tracking plane & 3.2 & 0.8 \\
Energy plane & 8.6 & 2.8 \\
Front-end electronics & 7.2 & 5.7 \\
Field cage & 8.2 & 1.7 \\ \midrule
Total & 37.2 & 12.6 \\
\bottomrule
\end{tabular}  
\end{center}
\end{table}
%%%%%%%%%%

In a second step we computed the background rejection factor achievable with the detector. Simulated events, after reconstruction, were accepted as a \bbonu\ candidate if
\begin{enumerate}
\item[(a)] they were reconstructed as a single track confined within the active volume;
\item[(b)] their energy fell in the region of interest, defined as $\pm 0.5$ FWHM around \Qbb; 
\item[(c)] the spatial pattern of energy deposition corresponded to that of a \bbonu\ track (\emph{blobs} in both ends).
\end{enumerate}
The achieved background rejection factor together with the selection efficiency for the signal are shown in table \ref{tab:RF}. As can be seen, the cuts suppress the radioactive background by more than 7 orders of magnitude.

%%%%%%%%%%
\begin{table}
\caption{Suppression of \BI\ events by the selection cuts.}
\label{tab:RF}
\begin{center}
\begin{tabular}{lccc}
\toprule
 & \multicolumn{3}{c}{Fraction of events} \\
Selection cut & \bbonu\ & \BI\ & \TL\ \\ \midrule 
Confined, single track & 0.48 & $6.0\times10^{-5}$ & $2.4 \times 10^{-3}$ \\
Energy ROI & 0.33 & $2.2\times10^{-6}$ & $1.9 \times 10^{-6}$ \\
Topology \bbonu\ & 0.25 & $1.9\times10^{-7}$ & $1.8 \times 10^{-7}$ \\
\bottomrule
\end{tabular}
\end{center}
\end{table}%
%%%%%%%%%%

Lastly, we multiply the gamma flux by the rejection factor to obtain the background rate (per unit of \bb\ isotope mass, energy and time) of the experiment:
\begin{eqnarray}
\nonumber b & = & \frac{(3.7\times10^{6})\cdot(1.9\times10^{-7})+(1.3\times10^6)\cdot(1.8\times10^{-7})}{100~{\rm kg}\cdot12.5~{\rm keV}}\\
&\simeq& 8 \times 10^{-4}\ \ckky.
\end{eqnarray}
If 50 extra kilograms of enriched xenon can be procured, the background rate would be even lower: $5. \times 10^{-4} \ckky$.

\section{Summary and outlook}\label{sec:Summary}
%%%
We have presented the technical design of the NEXT-100 detector, a high-pressure xenon gas time projection chamber that will search for \bbonu\ in \XE. The detector will use electroluminescence for the amplification of the ionization signal, and will have separate readout planes for tracking (an array of MPPCs) and calorimetry (an array of PMTs). Such a design provides both optimal energy resolution and event topological information for background rejection.

\begin{figure}
\centering
\includegraphics[width=10cm]{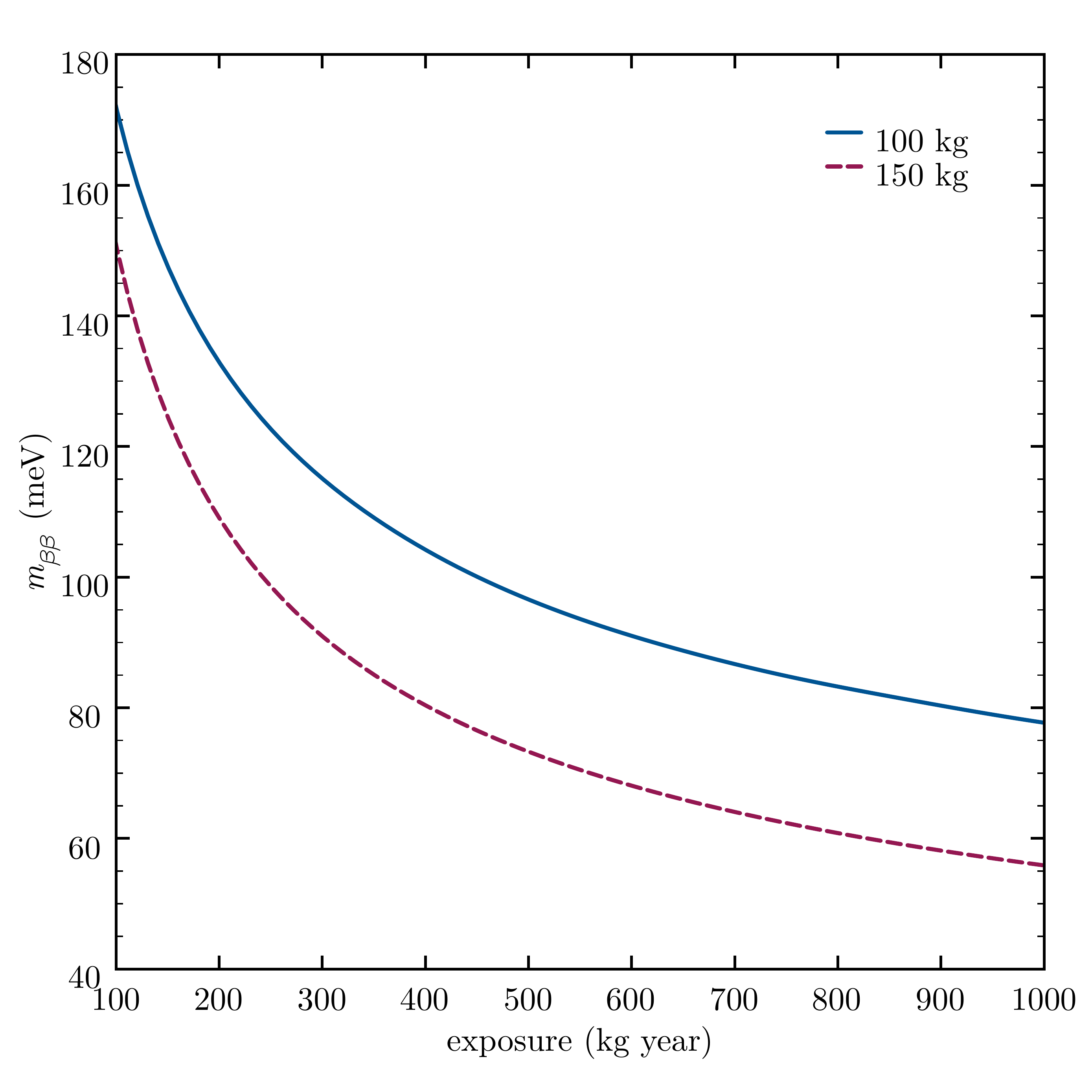}
\caption{Sensitivity (at 90\% CL) of NEXT-100 to the effective neutrino Majorana mass \mbb, computed following the method described in \cite{GomezCadenas:2010gs}. The solid, blue line corresponds to the baseline scenario where 100 kg of enriched xenon are used, whereas the dashed, red line shows the sensitivity of the detector with 150 kg of source mass.} \label{fig:sensi}
\end{figure}

The expected background rate is $8\times10^{-4}$ \ckky; this results in a sensitivity (see figure \ref{fig:sensi}), after 5 years of data-taking, of about $5.9\times10^{25}$ years or, in terms of \mbb, better than 100 meV.

NEXT-100 is approved for operation in the Laboratorio Subterr\'aneo de Canfranc (LSC), in Spain. The installation of shielding and ancillary systems will start in the second half of 2012. The assembly and commissioning of the detector is planned for early 2014.  

%%%%%%%%%%
%%%%%%%%%%

%%%%%%%%%%%%%%%%%%%%%%%%%%%%%%%%%%%%%%%%%%%%%%%%%%%%%%%%%%%%
\acknowledgments
The NEXT Collaboration acknowledges support from the following agencies and institutions:  the Spanish Ministerio de Econom\'ia y Competitividad under grants CONSOLIDER-Ingenio 2010 CSD2008-0037 (CUP), FPA2009-13697-C04-04 and RYC-2008-03169; the Portuguese Foundation for Science and Technology under grant PTDC/FIS/112272/2009 (``High Pressure Xenon Doped Mixtures for NEXT Collaboration''); the European Commission under the European Research Council Starting Grant ERC-2009-StG-240054 (T-REX) of the IDEAS program of the 7th EU Framework Program; and the Director, Office of Science, Office of Basic Energy Sciences, of the US Department of Energy under contract no.\ DE-AC02-05CH11231. J.~Renner (LBNL) acknowledges the support of a US DOE NNSA Stewardship Science Graduate Fellowship under contract no.\ DE-FC52-08NA28752. 

\bibliographystyle{JHEP}
\bibliography{references}

\providecommand{\href}[2]{#2}\begingroup\raggedright\begin{thebibliography}{10}

\bibitem{GomezCadenas:2011it}
J.~J. G\'omez-Cadenas, J.~Mart\'in-Albo, M.~Mezzetto, F.~Monrabal, and
  M.~Sorel, {\it The search for neutrinoless double beta decay},  {\em Riv.
  Nuovo Cim.} {\bf 35} (2012), no.~2 29--98,
  [\href{http://xxx.lanl.gov/abs/1109.5515}{{\tt arXiv:1109.5515}}].

\bibitem{Granena:2009it}
{\bf NEXT} Collaboration, F.~Gra\~nena {\em et~al.}, {\it {NEXT, a HPGXe TPC
  for neutrinoless double beta decay searches}},
  \href{http://xxx.lanl.gov/abs/0907.4054}{{\tt arXiv:0907.4054}}.

\bibitem{Alvarez:2011my}
{\bf NEXT} Collaboration, V.~\'Alvarez {\em et~al.}, {\it {The NEXT-100
  experiment for neutrinoless double beta decay searches (Conceptual Design
  Report)}},  \href{http://xxx.lanl.gov/abs/1106.3630}{{\tt arXiv:1106.3630}}.

\bibitem{Schechter:1981bd}
J.~Schechter and J.~W.~F. Valle, {\it {Neutrinoless Double beta Decay in
  SU(2)$\times$U(1) Theories}},  {\em Phys. Rev.} {\bf D25} (1982) 2951.

\bibitem{KlapdorKleingrothaus:2000sn}
H.~Klapdor-Kleingrothaus, A.~Dietz, L.~Baudis, G.~Heusser, I.~Krivosheina, {\em
  et~al.}, {\it {Latest results from the Heidelberg-Moscow double beta decay
  experiment}},  {\em Eur. Phys. J.} {\bf A12} (2001) 147--154,
  [\href{http://xxx.lanl.gov/abs/hep-ph/0103062}{{\tt hep-ph/0103062}}].

\bibitem{KlapdorKleingrothaus:2001ke}
H.~Klapdor-Kleingrothaus, A.~Dietz, H.~Harney, and I.~Krivosheina, {\it
  {Evidence for neutrinoless double beta decay}},  {\em Mod. Phys. Lett.} {\bf
  A16} (2001) 2409--2420, [\href{http://xxx.lanl.gov/abs/hep-ph/0201231}{{\tt
  hep-ph/0201231}}].

\bibitem{Aalseth:2002dt}
C.~Aalseth, I.~Avignone, F.T., A.~Barabash, F.~Boehm, R.~Brodzinski, {\em
  et~al.}, {\it {Comment on 'Evidence for neutrinoless double beta decay'}},
  {\em Mod. Phys. Lett.} {\bf A17} (2002) 1475--1478,
  [\href{http://xxx.lanl.gov/abs/hep-ex/0202018}{{\tt hep-ex/0202018}}].

\bibitem{Fukugita:1986hr}
M.~Fukugita and T.~Yanagida, {\it {Baryogenesis Without Grand Unification}},
  {\em Phys. Lett.} {\bf B174} (1986) 45.

\bibitem{Davidson:2008bu}
S.~Davidson, E.~Nardi, and Y.~Nir, {\it {Leptogenesis}},  {\em Phys. Rept.}
  {\bf 466} (2008) 105--177, [\href{http://xxx.lanl.gov/abs/0802.2962}{{\tt
  arXiv:0802.2962}}].

\bibitem{GomezCadenas:2010gs}
J.~J. G\'omez-Cadenas, J.~Mart\'in-Albo, M.~Sorel, P.~Ferrario, F.~Monrabal,
  {\em et~al.}, {\it {Sense and sensitivity of double beta decay experiments}},
   {\em JCAP} {\bf 1106} (2011) 007,
  [\href{http://xxx.lanl.gov/abs/1010.5112}{{\tt arXiv:1010.5112}}].

\bibitem{Nygren:2009zz}
D.~Nygren, {\it {High-pressure xenon gas electroluminescent TPC for 0-$\nu$
  $\beta\beta$-decay search}},  {\em Nucl. Instrum. Meth.} {\bf A603} (2009)
  337--348.

\bibitem{Ackerman:2011gz}
{\bf EXO} Collaboration, N.~Ackerman {\em et~al.}, {\it {Observation of
  Two-Neutrino Double-Beta Decay in $^{136}$Xe with EXO-200}},  {\em Phys. Rev.
  Lett.} {\bf 107} (2011) 212501,
  [\href{http://xxx.lanl.gov/abs/1108.4193}{{\tt arXiv:1108.4193}}].

\bibitem{Gando:2012kz}
{\bf KamLAND-Zen} Collaboration, A.~Gando {\em et~al.}, {\it Measurement of the
  double-beta decay half-life of {$^{136}$Xe in KamLAND-Zen}},
  \href{http://xxx.lanl.gov/abs/1201.4664}{{\tt arXiv:1201.4664}}.

\bibitem{Luscher:1998sd}
R.~Luscher, J.~Farine, F.~Boehm, J.~Busto, K.~Gabathuler, {\em et~al.}, {\it
  {Search for beta beta decay in Xe-136: New results from the Gotthard
  experiment}},  {\em Phys.Lett.} {\bf B434} (1998) 407--414.

\bibitem{Renner:2011}
A.~Goldschmidt, T.~Miller, D.~Nygren, J.~Renner, D.~Shuman, H.~Spieler, and
  J.~White, {\it {High-pressure xenon gas TPC for neutrino-less double-beta
  decay in \XE: Progress toward the goal of 1\% FWHM energy resolution}}, .
  {Submitted to the 2011 IEEE Nuclear Science Symposium}.

\bibitem{Aprile:2011ru}
{\bf XENON} Collaboration, E.~Aprile, K.~Arisaka, F.~Arneodo, A.~Askin,
  L.~Baudis, {\em et~al.}, {\it {Material screening and selection for
  XENON100}},  {\em Astropart.Phys.} {\bf 35} (2011) 43--49,
  [\href{http://xxx.lanl.gov/abs/1103.5831}{{\tt arXiv:1103.5831}}].

\bibitem{Boccone:2009kk}
{\bf ArDM} Collaboration, V.~Boccone {\em et~al.}, {\it {Development of
  wavelength shifter coated reflectors for the ArDM argon dark matter
  detector}},  {\em JINST} {\bf 4} (2009) P06001,
  [\href{http://xxx.lanl.gov/abs/0904.0246}{{\tt arXiv:0904.0246}}].

\bibitem{MPPC}
\href{http://jp.hamamatsu.com/products/sensor-ssd/4010/S10362-11-050P/index_en.html}{http://jp.hamamatsu.com/products/sensor-ssd/4010/S10362-11-050P/}.

\bibitem{Alvarez:2012ub}
{\bf NEXT} Collaboration, V.~\'Alvarez {\em et~al.}, {\it {SiPMs coated with
  TPB: coating protocol and characterization for NEXT}},  {\em JINST} {\bf 7}
  (2012) P02010, [\href{http://xxx.lanl.gov/abs/1201.2018}{{\tt
  arXiv:1201.2018}}].

\bibitem{Lung:2012pi}
K.~Lung, K.~Arisaka, A.~Bargetzi, P.~Beltrame, A.~Cahill, {\em et~al.}, {\it
  {Characterization of the Hamamatsu R11410-10 3-Inch Photomultiplier Tube for
  Dark Matter Direct Detection Experiments}},
  \href{http://xxx.lanl.gov/abs/1202.2628}{{\tt arXiv:1202.2628}}.

\bibitem{CanfrancFlux}
G.~Luz\'on {\em et~al.}, {\it {Characterization of the Canfranc Underground
  Laboratory: Status and future plans"}},  in {\em Proceedings of the 6th
  International Workshop on Identification of Dark Matter}, vol.~514,
  September, 2006.

\bibitem{ILI1}
\href{http://radiopurity.in2p3.fr/}{http://radiopurity.in2p3.fr/}.

\bibitem{BOC09}
V.~Boccone {\em et~al.}, {\it Development of wavelength shifter coated
  reflectors for the ardm argon dark matter detector'},  {\em JINST} {\bf 4}
  (2009) P06001, [\href{http://xxx.lanl.gov/abs/0904.0246}{{\tt
  arXiv:0904.0246}}].

\bibitem{SNOrp}
I.~Lawson and B.~Cleveland, {\it Low background counting at snolab},  in {\em
  Topical Workshop on Low Radioactivity Techniques LRT 2010}, pp.~68--77, AIP
  Conf. Proc. 1338, 2011.

\bibitem{BUD09}
D.~Budjas {\em et~al.}, {\it Gamma-ray spectrometry of ultra low levels of
  radioactivity within the material screening program for the gerda
  experiment},  {\em Applied Radiation and Isotopes} {\bf 67} (2009) 755.

\bibitem{FAH11}
C.~Faham, {\it Performance and radioactivity measurements of the pmts for the
  lux and lz dark matter experiments},  in {\em TIPP2011}, 2011.
\newblock \url{http://conferences.fnal.gov/tipp11/}.

\bibitem{LEO08}
S.~Leonard {\em et~al.}, {\it Systematic study of trace radioactive impurities
  in candidate construction materials for exo-200},  {\em NIMA} {\bf 591}
  (2008) 490.

\bibitem{NIS09}
S.~Nisi {\em et~al.}, {\it Comparison of inductively coupled mass spectrometry
  and ultra low-level gamma-ray spectroscopy for ultra low background material
  selection},  {\em Applied Radiation and Isotopes} {\bf 67} (2009) 828.

\end{thebibliography}\endgroup

\end{document}